\documentstyle[prc,aps,manuscript,12pt]{revtex}
\draft
\begin{document}

\title{Antisymmetrization in the Multicluster Dynamic
Model of Nuclei and the Nucleon Exchange Effects}
\author{R.A. Eramzhyan$^{1,2}$, G.G. Ryzhikh$^{1,3}$,
and Yu.M. Tchuvil'sky$^4$  \\
$^1$ Institute for
Nuclear Research of the Russian Academy of Sciences, Moscow
117312, Russia    \\
$^2$ Cyclotron Institute, Texas A$\&$M University, College Station,
TX 77843-3366,USA  \\
$^3$ Faculty of Science, Northern Territory University,
Darwin NT 0909, Australia \\
$^4$ Institute of
Nuclear Physics  of Moscow State University, Moscow
119899, Russia}

\maketitle

\begin{abstract}
A modified version of the Multicluster Dynamic Model of nuclei is
proposed to construct
completely antisymmetrized wave functions of multicluster systems.
An overlap kernel operator is introduced to renormalize the total wave
function after antisymmetrization between nucleons in different clusters.
A group-theoretical method is developed to analyze the role of the
exchange effects arising in the calculation of the various observables of
multicluster systems due to this antisymmetrization.

    The Antisymmetrized version of the Multicluster Dynamic Model is applied
to the six-nucleon systems treating them as $\alpha$-2N ones.
The static and dynamic
characteristics of the six-nucleon systems manifested in electron and
$\pi$-meson scattering,
muon capture, $\beta$-decay, pion photoproduction, etc., are calculated.
Significant progress is achieved in describing of variety of dynamic
observables of the six-nucleon systems as compared to the multicluster
dynamic model. In most cases calculated static and especially
dynamic characteristics are in a good agreement with the
experimental data.
\end{abstract}

\vspace{1cm}


\vspace{1cm}


\section{Introduction}

     To describe the structure of light nuclei (A$>$4) in great
detail very sophisticated approaches are now being elaborated. Recently,
Green's Function Monte Carlo calculations
have been realized \cite{gfmc} for nuclei with A$\le 6$ starting from
the realistic nucleon-nucleon interaction. At the same time
the stochastic variational method (this method was first applied
in calculations of nuclear structure within the multicluster
dynamic model (MDM) in Ref.\cite{KK})
has been extended to the systems with A$\le 10$ treating all
nucleons variables on the same ground
(see, for example, Ref.\cite{svm}). Before that the K-harmonic
method \cite{K-harm,gorbatov} and the extended
shell model \cite{sh-mod,sh2} have been used to learn more about
the structure of the 1p-shell nuclei. However, practical applications of
all these methods is not an easy task.
That is why there are not many observables analyzed with
these newly developed
methods. For this reason they are unable to displace the traditional
approaches, considering the fact that the latter are improved
as well.

     The Microscopic Cluster Model (MCM) still continues to play a
role as a very successful approach well suited for description  of
the structure of light nuclei. This model in its development has passed
many stages,
starting from its simplest realization within the framework of the
single
channel Resonating Group Method (RGM) \cite{1-rgm} and coming finally
to the many-channel multicluster MCM \cite{lang,varga1,varga2,svm}.

This model  proceeds  from  the  tendency of the nucleons
to form  clusters inside the light nuclei due to
their strong binding. In the MCM the A-nucleon wave  function
is approximated by a superposition of the  antisymmetrized basis functions
in which nucleons have been grouped into some clusters --- $\alpha$, $^3$He,
$^3$H, $d$ etc. The  cluster
internal wave functions are supposed to be known from the beginning
but  the relative motion function is obtained from the solution of
the dynamic equation (or also from
the system of coupled equations)
derived straightforwardly from the initial A-nucleon
problem.

 This approach in its final version  enables us to incorporate into the
calculation various multicluster  configurations,  to  take
into  account some effects of the cluster  excitations  and  their
rearrangement,  to treat nucleons on the same ground as clusters, etc.
Finally a consistent and successful description  of bound,
resonant and scattering states of many nuclei
\cite{lang,varga1,varga2,wt} has been achieved.

At the same time extensive  applications  of  the  MCM  have revealed
some deficiencies  which show  up  mainly  in the case  when the
short-range  correlations  among  nucleons  appear to be important.
Refined realization of the model requires that
more realistic nucleon-nucleon interaction should be used and,
consistently, more realistic wave functions of clusters should be
involved in the calculation.
But such a program is extremely difficult to realize in practice.
Indeed the cluster-cluster potential which  finally  arises  in
the MCM comes about from the folding of the NN  potential  over
the intrinsic wave functions of  the  constituent  clusters.  This
potential appears to be  very  complicated,  highly  nonlocal  and
energy dependent due to the exchange effects  and the composite
nature of clusters.  If one starts
from the cluster internal wave functions  with both short
range  and tensor NN  correlations  and uses the
realistic NN forces with the noncentral  components, then one arrives to
such a complicated cluster-cluster potential that it becomes almost
useless
for practical applications. That is why a simple intrinsic
wave function of cluster and a simplified effective NN  forces
are usually used in practice. Of course in this case this approach
looses its "pure" microscopic character and it becomes model dependent.

       On the other hand, when one starts from  the NN potential
with a strong short-range repulsion and uses
noncorrelated wave functions of
clusters then the results of the calculation become highly
dependent on the number and the type of the cluster
channels taken    into account         \cite{fuji,kruppa}.
In spite of this fact, many calculations  are still
undertaken in this approximation with forces
like the Hasegawa-Nagata or the
Eikemeier-Hackenbroich one (see, for example,
Refs.~\cite{kajino,hofmann}).

There is another problem which is a principal
one for the MCM. Derived from the more or less realistic  NN potential,
the cluster-cluster potential
of the MCM does not reproduce accurately the phase shifts for free
clusters. This problem was studied thoroughly in the case of
the simplest cluster system, namely for the $\alpha$-N
(see Ref.~\cite{efros}). Here the N-$\alpha$ potential was
constructed starting from a
realistic NN potential and a very complicated
$\alpha$-particle wave function.
It turned out that even in this simple case, in order to
reproduce  the  experimental  $\alpha$-N  phase-shifts
the starting realistic
nucleon-nucleon  potential had to be modified significantly.
It means that in the MCM, a phenomenological (effective) NN-potentials
should be used rather than the realistic ones.

     At the same time a microscopically inspired
Potential  Cluster  Model, which
is also called the  Multicluster Dynamic Model --- MDM, has been
also  applied successfully for the calculation of the structure
of light nuclei.
Formulation of  the  MDM  is given, for example, in Ref.~\cite{clust}.
In this approach the relative motion function of clusters is obtained
as a solution of the $k$-cluster dynamic equations
with the cluster-cluster potentials derived  directly
from  the phase-shifts analysis. Thus, in  the MDM  the  on-shell
behavior  of  the
interaction amplitudes are taken from the outset  with
higher precision as compared to the  MCM.

In the MDM there is no problem  at all
with the intrinsic  cluster wave functions due to the  factorization
of  the matrix elements squared for the nuclear observables.
Due to this factorization, it is not the intrinsic cluster wave
functions but rather the magnitudes of their
observables (that are directly taken from experiments) which are used here.
In this way many  observables  of the clusterized
nuclei were  obtained in a good agreement with the
experimental data (see, for example, Refs.~\cite{fun92,be9,zhukov}).
Another attractive point of the MDM is its much more simple
formalism as compared to the MCM.
These simplifications come from the approximate treatment  of  the
Pauli-principle  for  the nucleons from different clusters.

However, there are some  observables  where just the
nucleon  exchanges  between different clusters   are   of   primarily
importance. That is why we have modified the MDM to
construct finally completely
antisymmetrized wave functions. This modified version of the model was
called the AMDM --- the Antisymmetrized version of the MDM.
It was shown in Ref.~\cite{amdm} that by using the AMDM we extend the list
of well reproduced observables as compared to the  MDM.  In
particular, it was possible to describe all longitudinal and
transversal elastic and inelastic electromagnetic form factors of $^6$Li
simultaneously for a broad
interval of the momentum transfer. At the  same time the AMDM
formalism is still more simple and transparent than the MCM one.

     The method of construction  of the totally
antisymmetric wave function from the MDM function which had been used
in our earlier  works
\cite{amdm,physlet}
has some limitations. The fact is  that  after antisymmetrization
of the MDM functions between nucleons from the different clusters,
the set of total wave functions obtained loses
the orthonormalization property. First of all one  has  to
restore the total normalization of the wave functions. In
Ref.~\cite{physlet}
it was done by the simplest way of multiplying the wave function
by a constant. This is a common procedure for  the  shell  model
approach where the number of oscillator quanta for each wave function
is fixed. However, in general case such a procedure is not
quite correct. The reason is that the constant renormalization
($C$-renormalization as we will abbreviate it further) does not restore
the orthogonality between different antisymmetrized wave functions
with the same total quantum numbers.
Moreover, this $C$-renormalization changes
the asymptotic  part of the relative motion wave function  where  the
antisymmetrization  must have no influence at all.

        Experience  accumulated
in the MCM suggests that the renormalization with the integral
exchange  kernel
($K$-renormalization), which  appears primarily in the RGM, will
allow us to solve this problem in the general case as well.
It is very important that there is a kernel renormalization that
allows
to determine the class of observables which are exactly conserved under
the antisymmetrization \cite{solov}.
Some aspects of the renormalization
problem have been discussed already in Ref.~\cite{izvest} for $^6$Li
which was treated as a three-cluster system. In this paper we present
new precisely renormalized version of the AMDM and its
application to many observables of the six-nucleon systems.
The six-nucleon system is particularly singled out in nuclear physics
and has been studied in many papers as a three-nucleon system.
Six nucleons in nuclei exhibit quite properly the typical properties of
all light nuclei, such as the decisive role of the Pauli principle,
the substantial spin-orbit splitting of levels, a significant greater
role of the interaction of valent nucleons with core compared with the
interaction among them, the important contribution of the short-range
correlations of nucleons, etc. These circumstances, as well as numerous
experimental data obtained for six-nucleon systems including
the strong, electromagnetic and weak interaction processes permit
their theoretical studies to be regarded as a theoretical laboratory
for nuclear physics in a broad sense allowing to test new theoretical
approaches.

     This paper is organized in the following way.
In  Section II the formalism of the AMDM is briefly described and
a group-theoretical analysis of matrix elements for various
observables is performed.
In Section III many observables of the A=6  nuclei treated as a
three-cluster $\alpha$-2N
systems are analyzed within the framework of both versions of the AMDM
--- with constant and kernel renormalization. All these results
are compared
to those obtained within the MDM. Besides the traditional observables ---
static properties of ground state, beta-decay and electron scattering ---
we are discussing muon capture, pion scattering and pion photoproduction.
All these reactions proceed via similar nuclear transition operators.
Studying all of these transitions does not simply lead to a large
amount of redundant information, but, rather, it generally leads
to complementary information. So the most informative approach
for studying the nuclear structure is to carry out a simultaneous
study of all of them.
We conclude in Section IV.

\section{General properties of the AMDM}

\subsection{Construction of the MDM wave function}

     The AMDM is essentially based on the MDM. Therefore
it makes sense to  run briefly through the basic ingredients of the MDM.

    Let  $A$ identical fermions  (nucleons in our case, but they could
be quarks,  etc.) be distributed over  $k<A$ clusters. Individual
nonclusterized (valent) fermions, if they appear in this case,
are treated as  clusters as well.

     In the MDM, for a system made up  of the $k$ clusters
the wave function, with given total momentum $J$,
total isospin $T$ and their projections $M$ and $M_T$, is defined as
\begin{equation}
\Psi^{\rm MDM}_{JM}(\{\vec\xi^{tot}\})=
\sum_{\{\tilde t\}\{\tilde j\}\{ l\}}
\left\{\left\{\Phi^{int}\right\}^k_{TM_T,\, J_cM_c,}
\Psi_{LM_L}(\{\vec\xi\})
\right\}_{JM} ,                                      \label{fun}
\end{equation}
where $J_c$ is the channel spin,
$\{\vec\xi^{tot}\}$ denotes the complete set of the intercluster Jacoby
coordinates.
$\Psi_{LM_L}(\{\vec\xi\})$ is the function of the cluster-cluster
relative motion with the total angular momentum $L$.
It depends on the corresponding set of the
Jacoby coordinates $\{\vec\xi\}$. Each coordinate
$\vec\xi_j$
is divided into  angular
and spatial parts: $\vec\xi_j\ = (\hat\xi_j,\xi_j)$.

We first define some notations to be used in this Section.
Let's $x$ to be  any momentum ($l$ --- angular momentum, $s$ --- spin,
$\vec j = \vec l + \vec s$ ---
total momentum, $t$ ---isospin) and $m_x$ its projection, then the
symbol $\{...\}^k_{xm_x}$ will denote the coupling scheme of
the individual momenta
$x_i$
and their projections $m_{x_i}$ to the intermediate
$x_{12},...x_{12...k}$ momenta and the total, $x$, momentum and
its projection:

\begin{equation}
\{...\}^k_{xm_x}=\left\{\{\{x_1x_2\}_{x_{12}}x_3\}_
{x_{123}}...x_k\right\}_{xm_x} .                        \label{coupl}
\end{equation}
Correspondingly, the symbol with double down index is
defined as:
\begin{equation}
\{...\}^k_{xm_x,\, ym_y}=
\{...\}^k_{xm_x}\{...\}^k_{ym_y} \ .
\end{equation}

The following abbreviations are used to list the  intermediate momenta
in eq.~(\ref{fun}):
\begin{eqnarray}
\{\tilde j\}&=&(j_{12},j_{123},...,j_{12...k},J_c) , \nonumber \\
\{\tilde t\}&=&(t_{12},t_{123},...,t_{12...k},T) , \nonumber \\
\{ l\}&=&
(l_1,l_2,...,l_{k-1},l_{12},l_{123},...,l_{12...k-1},L)  ,
\end{eqnarray}
where $l_i$ is the orbital momentum
conjugated with the $i$-th Jacoby coordinate of
the relative motion; $L$ is the total angular momentum
of the system, $J_c$ is the spin of the channel. The total momentum
$J$ equals to the sum
$\vec J = \vec J_c + \vec L$.
Finally the complete set of quantum numbers $\{ \omega\}$
which characterizes any channel of
the system  is defined as
\begin{equation}
\{ \omega\}=(\{\tilde j\},\{\tilde t\},\{l\},J) .
\end{equation}

According to the MDM, the internal
wave function of the clusters
$\Phi^{int}$ is not modified in the nucleus and it is built up from
the wave functions of free clusters in their ground states:
\begin{equation}
\Phi^{int}=\prod_{i=1}^{k}
\phi^i_{j_im_i,t_im_{t_i}}
(\{\vec\xi^{int}_i\}) ,
\end{equation}
where $\{\vec\xi^{int}_i\}$ is the set of the internal Jacoby coordinates
for the $i$-th cluster and  $j_i$ and $t_i$ are their total momentum
and isospin with projections $m_i$ and $m_{t_i}$, respectively.
The role of the cluster polarization inside a nucleus has been
studied recently
within the MCM. It was shown in Refs.~\cite{fuji,kruppa} that
the $\alpha$-cluster is  polarized very weakly inside light nuclei.
Therefore, the
assumption that the $\alpha$-cluster can be considered
as a free $\alpha$-particle is rather good in any multicluster approach.

The wave function of the cluster-cluster relative motion
$\Psi_{LM_L}(\{\vec\xi\})$ is decomposed  into the angular
$Y_{\{ l\}M_l}(\hat\xi)$ and the radial
$F_{\{ \omega\}}(\xi)$ parts.
The angular part of the relative wave function  can be written as:
\begin{equation}
Y_{\{ l\}M_l}(\hat\xi) =\left\{\prod_{i=1}^{k-1} Y_{l_im_{l_i}}
(\hat\xi_i)\right\}^{k-1}_{LM_L} \ .
\end{equation}

It is very convenient to present the radial wave function of the relative
motion in terms of the superposition of gaussians (it was demonstrated in
many papers, see for example Ref.~\cite{fun92}) as follows:
\begin{equation}
F_{\{ \omega\}}(\xi)=\sum_{\nu}C^{\{\omega\}}_\nu
\prod_{i=1}^{k-1} |\xi_i|^{l_i}\exp(-\alpha^{\{\omega\}}_{\nu i}
\xi_i^2)\ ,
\end{equation}
where $C^{\{\omega\}}_\nu$ and $\alpha^{\{\omega\}}_{\nu i}$ are the
expansion coefficients. The set of these
coefficients together with the set of the "channel" quantum numbers
$\{\omega\}$  determine completely the MDM wave function.
The variational method for solution of the $k$-cluster problem is well
suited for finding these coefficients. As it was shown in
Refs.~\cite{fun92,be9},
such an approach allows to obtain a very accurate solution of
the nuclear few-body problem. It is the gaussian parameterization
of the MDM  wave function that makes it possible
to get analytical expression
for most matrix elements both in the MDM and in the AMDM
(see, for example, Refs.~\cite{amdm,physlet,prepr}).

As it was shown by
Saito \cite{saito1} that sophisticated nonlocal energy-dependent
"microscopic" cluster-cluster forces appearing in the RGM can
be successfully approximated  by a simple local $E$-independent ones
if they are supplemented by the orthogonality requirement based on the
Pauli principle. In such an approach, both the phase shifts
and the nodal structure  of the RGM wave functions
can be quite well described in a very simple way.
This is the basic approximation for the
Orthogonality Condition Model (OCM) \cite{saito1,saito2}
and, in fact, for any MDM.
In the  version of the
MDM \cite{fun92} to be used throughout this paper, the Pauli-forbidden
components are excluded from the solution of the MDM
dynamic equations by means of the pseudopotential technique
\cite{fun92,psepot}.
As we already stated before, such "microscopically inspired"
intercluster forces have even some advantage
compared to the "true microscopic" ones. This advantage consists
in more accurate description of the low-energy cluster-cluster phase
shifts and, therefore, the most important on-shell properties of the
cluster-cluster forces.

However, exclusion solely of the forbidden components
does not account for
all the consequences of the Pauli principle for
the multicluster system. Exchanges between nucleons
in the different clusters are completely omitted in the MDM.
That is why  some types of observables can not be explained in the
framework of this model.
Therefore we came to recognize that the totally Antisymmetrized
MDM (the AMDM) should be used in general case \cite{amdm,physlet}.
Notice that the AMDM is, in fact, a modification and generalization
of the OCM.

\subsection{ Wave functions of the AMDM}

     In the initial version of  the  AMDM \cite{amdm} named here as
AMDM$_{\rm C}$
the  total  wave function was constructed by means of direct
antisymmetrization of the MDM wave function:
\begin{equation}
\Psi_A=\frac{1}{Q}\hat{A}\Psi^{MDM}\ .  \label{amdmc}
\end{equation}
As we have noted above, in this case we use a constant $Q$ to
normalize the total function $\Psi_A$. That is why we call
this version
of the model as the AMDM$_{\rm C}$ --- Antisymmetrized Multicluster
Dynamic Model with the renormalization to the constant.
$\hat A$ in eq.(\ref{amdmc}) is an antisymmetrizer:
\begin{equation}
\hat A = \Omega^{-1}\left
(1+\sum _P(-1)^p\hat {P}\right )  \ ,
\end{equation}
where the normalizing factor $\Omega$ has the following form:
\begin{equation}
\Omega = \left (\frac {A!}{\prod _{i=1}^{k} A_i!} \right )^{\frac
{1}{2}}.
\end{equation}
Here $A_i$ is the mass of the individual cluster  $i$,
$\hat {P}$ is
the operator permutating nucleons between different clusters and
$p$ is the parity of such a permutation.

     In a new version of the AMDM, named the AMDM$_{\rm K}$, the overlap
kernel operator $\hat K$ is introduced to renormalize
the total wave function after antisymmetrization:

\begin{equation}
\Psi_A=\hat A\sum_{\{\tilde t\}\{\tilde j\}\{ l\}}
\left\{\left\{\Phi^{int}\right\}^k_{TM_T,\, J_cM_c}\cdot\hat K^{-1/2}
\Psi_{LM_L}(\{\hat\xi\})\right\}_{JM} .    \label{amdmk}
\end{equation}
This operator $\hat K$  acts in the Hilbert space of
the $(3k-3)$-dimensional orbital
functions of the $k$-cluster system (cluster subspace).
It acts on the function in the following way:
\begin{equation}
\hat K \Psi(\{\vec\xi '\} ) = \int d\{\vec\xi "\}
K(\{ \vec\xi ',\vec\xi "\} ) \Psi(\{\vec\xi "\} ) .
\end{equation}

The overlap kernel $K(\{ \vec\xi ',\vec\xi "\} )$ is defined as
a projection of the antisymmetrizer onto the cluster subspace of
$A$-nucleon variables:
\begin{equation}
K(\{ \vec\xi ',\vec\xi "\} ) = \Omega\cdot\left\langle
\left\{\Phi^{int}\right\}^k_{S,T}
\prod_{i=1}^{k-1}\delta (\vec \xi_i-\vec\xi '_i)
\mid\hat {A}\mid
\left\{\Phi^{int}\right\}^k_{S,T}
\prod_{i=1}^{k-1}\delta (\vec \xi _i-\vec\xi "_i)
\right\rangle .                                              \label{ker}
\end{equation}

The eigenfunctions $\Phi_\nu(\{\vec\xi\})$ and the eigenvalues
$\varepsilon_\nu$
of this kernel satisfy the usual equation:

\begin{equation}
\hat K \Phi_\nu(\{\vec\xi\}) = \varepsilon_\nu \Phi_\nu(\{\vec\xi\}) ,
                                                   \label{eigen}
\end{equation}
and
\begin{equation}
\hat K^{-1/2} \Phi_\nu(\{\vec\xi\}) =
\varepsilon^{-1/2}_\nu \Phi_\nu(\{\vec\xi\}) ,
\end{equation}
for
\begin{equation}
\varepsilon_\nu \neq 0 .
\end{equation}

In the MDM, any nucleus is usually considered as a system
composed of some number of magic clusters ($\alpha$-particles,
$^{16}$O, etc.) in their ground states and $n$-valent nucleons.
In this Section, to make the group-theoretical analysis more transparent
we restrict the general case by the following two conditions:
\begin{enumerate}
\item The magic clusters are consider to be the SU(3) and SU(4) scalars.
So, in their wave functions we neglect all components
violating these symmetries.
If the system is built up from different clusters
($\alpha$-particles
and $^{16}O$, for example) we will neglect also the difference in
their oscillator parameters $\bar h\omega$ in order to conserve the
SU(3) scalar character of $\left\{\Phi^{int}\right\}^k_{S,T}$.
\item The valent group of $n$ nucleons is considered either as one light
cluster with three or less nucleons (with the common oscillator parameter
$\omega$)
or as nonclusterized nucleonic system if $n\leq 2$.
\end{enumerate}
These conditions lead to that the total spin $S=J_c$ and isospin $T$
are uniquely determined by the Young scheme [f] and Yamanouchi
symbol $(r)$. These quantum numbers are precisely those that characterize
the MDM functions.

As a consequence of the SU(3)-symmetry, the
eigenfunctions of the operator $\hat{K}$ are the many-body oscillator
functions with a definite number of oscillator quanta $N$ and a
fixed value of the
Elliott symbols $(\lambda \mu)$.  Since
the same Elliott symbol can characterize several eigenfunctions,
an additional quantum number $i$ is introduced to distinguish among them.
The eigenfunctions of the operator $\hat K$ are characterized by
the total angular momentum $L$.
It is interesting to note that its eigenvalues
do not depend on $L$ at all. Finally
for a  complete definition of eq.(\ref{eigen})
we specify a symbol $\nu$ as:

\begin{equation}
\nu=(S,T,L,N,(\lambda\mu),i) .
\end{equation}
According to these  restrictions  the internal part of the MDM
function can be written down in the factorized form:

\begin{equation}
\left\{\Phi^{int}\right\}^k_{J_cM_c,\, TM_T}=
\Phi_{int}\cdot\left\{\prod_{i=1}^{k}
\chi_{s_it_i}\right\}^k_{SM_S,\, TM_T},
\end{equation}
where $\Phi_{int}$ is a scalar spatial function and the rest is
the spin-isospin one. To simplify the reading of the subsequent
equations we will
use the following notations for the spin-isospin function:
\begin{equation}
\chi_{ST}=\chi_{[f](r)}=
\left\{\prod_{i=1}^{k}\chi_{s_it_i}\right\}^k_{SM_S,\, TM_T}.
\end{equation}

Before concluding this subsection
we note that these restrictions  are used only
in Section II for the sake of
the formal analysis of the antisymmetrization effects.
As to the numerical calculations in Section II, more complicated
wave functions were actually used.
On the other hand, it appeared that the results of
the formal analysis are valid even in the
case when the conditions 1 and 2 are not fulfilled exactly.

\subsection{Formal analysis of the matrix elements
within the framework of the AMDM}

Before starting any discussion of the results of the calculations
within the framework of the AMDM
it is very instructive to make some classification of
the nuclear matrix elements.
For this reason let us expand the relative motion
functions of the MDM onto the eigenfunctions $\Phi_\nu$
of the overlap kernel:
\begin{equation}
\Psi(\{\vec\xi\})=\sum_{\nu}b_\nu \Phi_\nu(\{\vec\xi\}). \label{expan}
\end{equation}

Generally, the wave function in the multicluster approach can contain
some admixture of the Pauli-forbidden
components  with  $\varepsilon_\nu= 0$. After antisymmetrization these
components have to disappear. To simplify the subsequent expressions,
it is assumed below that such forbidden
components have been  eliminated from the beginning (i.e. $b_\nu=0$
 if $\varepsilon_\nu=0$).
So one can write down that:

\[ \langle \Phi \mid \hat {O}\mid \Phi '\rangle = \sum
_{\nu_1\nu_2}b_{\nu_1}b_{\nu_2}\varepsilon
^{-\frac {1}{2}}_{\nu_1}\varepsilon
^{-\frac {1}{2}}_{\nu_2}\cdot \]
\begin{equation}
\langle \hat {A}\left \{ \Phi_{int}\Phi _{\nu_1}(\{
\vec\xi _i\} )\chi _{[f](r)}\right \} \mid \hat {O}\mid \hat {A}\left \{
\Phi_{int}\Phi _{\nu_2}(\{ \vec\xi _i\} )
\chi '_{[f'](r')}\right \} \rangle .       \label{oper}
\end{equation}

     For any operator $\hat{O}$ which is symmetric over the nucleon 
permutations one can replace an antisymmetrizer, say in the left hand
side of eq.~(\ref{oper}), by the factor $\Omega$.
Let us then add to this expression the sum over the
complete set of the nonantisymmetrized wave functions:
\begin{equation}
1 = \sum _{(j)\nu[f"](r")}\mid \Phi^{(j)}_{int}\Phi _\nu(\{ \vec\xi _i\}
)\chi
_{[f"](r")}\rangle \langle \Phi ^{(j)}_{int}\Phi _\nu(\{ \vec\xi _i\}
)\chi
_{[f"](r")}\rangle ,              \label{unity}
\end{equation}
where the sum over $(j)$ includes all orbital excitations
of the constituent clusters. Because
the antisymmetrizer does not change
the values of the total spin S, isospin T and Young scheme [f] of
the system one arrives to the following expression for the matrix elements:
\[ \langle \Phi \mid \hat {O}\mid \Phi '\rangle = \Omega \sum
_{\nu_1\nu_2\nu(j)\nu(r")}b_{\nu_1}b_{\nu_2}\varepsilon ^{-\frac
{1}{2}}_{\nu_1}\varepsilon
^{-\frac {1}{2}}_{\nu_2}\langle \Phi ^{(0)}_{int}\Phi _{\nu_1}(\{
\vec\xi _i\} )\chi _{[f](r)}\mid \hat {O}\mid \Phi ^{(j)}
_{int}\Phi _\nu(\{ \vec\xi _i\} )\chi "_{[f'](r")}\rangle \cdot \]
\begin{equation}
\langle \Phi ^{(j)}_{int}\Phi _\nu(\{
\vec\xi _i\} )\chi" _{[f'](r")}\mid \hat {A}\mid \Phi ^{(0)}
_{int}\Phi _{\nu_2}(\{ \vec\xi _i\} )\chi '_{[f'](r')}\rangle .
                                                      \label{oper2}
\end{equation}
Here the index $(j)=(0)$ labels the intrinsic orbital functions
of clusters in their ground states.

    From eq.~(\ref{oper2}) it follows that if the operator $\hat O$
is diagonal in the space of the orbital functions
$\Phi ^{(j)}_{int}\Phi _{\nu}(\{\vec\xi _i\})$ and over the
Yamanouchi symbols $(r)$
then the matrix elements of the
MDM and AMDM are equal to each other because in this case $(j)=(0)$
only and
according to expressions (\ref{amdmk}), (\ref{ker}) and (\ref{eigen})
the matrix element of the antisymmetrizer reduces to
$\varepsilon_{\nu_2}\delta_{\nu \nu_2}$. Such a situation occurs when:
\begin{enumerate}
\item An operator is of pure spin-isospin type.
\item An operator is of particular spatial type being a function of
the Casimir operators of the SU(3) group and its subgroups.
\end{enumerate}
Indeed in the last case this operator  is diagonal because
the wave functions in
(\ref{oper2})  have a definite SU(3) symmetry, and the Elliott symbol
($\lambda\mu$) is
a good quantum
number for the system.

It is interesting to analyze the difference between the matrix elements
of the full tensor
operators and the matrix elements of the pure spatial ones.
For this purpose
let us use the following representation of the
wave function:
\begin{equation}
\hat {A} \mid  \Phi^{(0)}_{int}
\Phi _{\nu_2}(\{ \vec\xi _i\} )\chi_{[f](r)}\rangle  =
\frac {\Omega }{n_f}\sum_{r}\hat {C}^{[f]}_{rr}\mid \Phi
^{(0)}_{int}\Phi _{\nu_2} \chi _{[f](r)}\rangle,          \label{trans}
\end{equation}
where $\hat {C}^{[f]}_{rr}$  is the Young operator
which projects the wave function onto the wave function with
the definite Young scheme $[f]$ and Yamanouchi symbol $(r)$
($n_f$ is a dimension of the corresponding representation):
\begin{equation}
\hat {C}^{[f]}_{rr} = \frac {n_f}{A!}\sum
_{P}D^{[f]}_{rr}(P)\hat {P} \ .
\end{equation}
Here $D^{[f]}_{rr}(P)$ is the matrix of the corresponding
 representation  of the permutation group.

After substitution of eq.~(\ref{trans}) into eq.~(\ref{oper2}),
one arrives at the following expression:
\[
\langle \Phi \mid \hat {O}\mid \Phi '\rangle = \sum
_{\nu_1\nu_2\nu(j)(r)}b_{\nu_1}b_{\nu_2}
\langle \Phi ^{(0)}_{int}\Phi
_{\nu_1}(\{\vec\xi _i\} )\chi _{[f](r_0)}\mid \hat {O}\mid \Phi
^{(j)}_{int}\Phi _{\nu}(\{ \vec\xi _i\} )\chi '_{[f'](r)}\rangle
\]
\begin{equation}
Q^{-1}_{\nu_1}Q^{-1}_{\nu_2}
\langle \Phi ^{(j)}_{int}\Phi _{\nu}(\{ \vec\xi _i\} )\mid
\hat C^{[f']}_{r r}\mid \Phi
^{(0)}_{int}\Phi _{\nu_2} (\{ \vec\xi _i\} )\rangle,        \label{26}
\end{equation}
where
\begin{equation}
Q_\nu = \langle \Phi ^{(0)}_{int}\Phi _{\nu}(\{ \vec\xi _i\} )
\mid \hat C^{[f']}_{r r}
\mid \Phi
^{(0)}_{int}\Phi _{\nu} (\{ \vec\xi _i\} )\rangle^{\frac{1}{2}} =
\Omega^{-1}(n_f\varepsilon _\nu)^{\frac{1}{2}} .
\end{equation}
$Q_\nu$ is the normalization factor of the wave function
 $ C^{[f']}_{r r}\mid \Phi
^{0}_{int}\Phi _{\nu} (\{ \vec\xi _i\} )\rangle$.

Now we are ready to establish some relation between the
renormalization factor Q in eq.(\ref{amdmc}), which have appeared in
the AMDM$_{\rm C}$  and the eigenvalues of the
overlap kernel $\varepsilon_\nu$:
\begin{equation}
Q^2 = \frac{n_f}{\Omega}\sum _{\nu} (b_{\nu})^2 \varepsilon_{\nu},         
                                                               \label{Q}
\end{equation}

Let us compare now matrix elements of a pure spatial operator
with matrix
elements of a full tensor one. It follows from eq.~(\ref{amdmc})
that in the latter case only the terms with
the Yamanouchi symbols $(r)\neq(r_0)$ contribute to the AMDM matrix
element, contrary to the case of a pure spatial operator.
Evidently such terms are not
present in the MDM case independently of the tensor structure of any
operator. That is why the matrix
elements of a full tensor operator are expected to be modified
much more than the matrix elements of a pure
spatial one when the exchange terms are taken into account.

Eq.(\ref{26}) allows to establish some properties
of matrix elements appearing in various reactions as well.
To demonstrate this let us
consider, for example, the cluster knock out reaction at large momentum
transfer, though it is
beyond the scope of this paper. Nevertheless, it is instructive to
say some words about this case.
The amplitude of the corresponding process is proportional to the
matrix element given by the expression (\ref{26}).
As follows from this expression  not only the
$(j)=(0)$ term, but also the terms with $(j)\neq(0)$
contribute to the amplitude of this reaction. This fact was known
in the theory of cluster knock out reactions for the special case
of oscillator shell model wave functions. These $(j)\neq (0)$ terms
correspond to cluster deexcitations according to Ref.\cite{neud11}.
Expression (\ref{26}) demonstrates
that the cluster deexcitation terms appear in the amplitudes of the
cluster knock out reactions and similar processes in the general case
as well. Despite the fact that in the case of cluster knock out one
deals with the
final state wave function which was named as an asymptotic one (because
in the corresponding wave function  components with the
eigenvalues $\varepsilon_\nu\simeq 1$ dominate in the expansion
(\ref{expan}))
the sum over $(j)$ in (\ref{26}) is not reduced to the trivial term
with $(j)=(0)$.
To demonstrate this let us compare two expressions. In the first one
the antisymmetrizer is removed from the
right hand side of eq.(\ref{oper}), in the second one ---
from the left hand side.
It is easy to show that they will be equal to each other if the
following relation holds

\begin{equation}
Q_{\nu_1}=Q_{\nu_2}\ (\varepsilon_{\nu_1}=\varepsilon_{\nu_2})
                                                       \label{cond}
\end{equation}
for all terms with $b_\nu\neq 0$. However, this is not the case
due to the contribution from the  terms with $\varepsilon_\nu\neq 1$.
So, the transition amplitudes from the bound cluster state, even to the
asymptotic one, contain the deexcitation terms. In most calculations
these terms were not taken into account at all.
Some examples of the importance of taking into account this type of
exchange terms are given in Refs.\cite{neud11,neud12}.

Finally, as follows from our analysis, all operators can be subdivided
into three groups
according to the magnitude of the exchange effects in their matrix
elements:
\begin{enumerate}
\item  Full tensor operators. Their  matrix elements are
affected maximally  when going from the MDM to the AMDM$_{\rm K}$.
\item  Pure spatial operators. Their matrix elements are less sensitive
to the exchange effects as compared to the first case. If the
Pauli forbidden components in the MDM function are eliminated
in a thorough way then
the exchange effects  usually become rather small.
\item  Operators whose matrix elements strictly conserved when going from
the MDM to the AMDM$_{\rm K}$.
They are either pure spin-isospin operators or pure
spatial operators being  some functions of Casimir operators of SU(3)
group.
\end{enumerate}

In most cases the nuclear observables are associated not with a single
tensor operator but with their sum.
For example, an operator of the  magnetic moment has
two terms. One comes from the magnetization current and the other from the
convective one. If the total angular momentum of the system equals
to zero then the contribution of the last term vanishes.
More generally, if the spin and isospin of the system under
discussion are equal to zero both in the initial and in the final states,
then any operator of physical observable appears as an effective
spatial operator. In such a case the nucleon exchange effects should
not be too large. On the other hand operators of some nuclear
observables (for example the operator of M1 transition) at low momentum
transfer
can be considered with a good approximation as an effective
spin-isospin one. In this limit the exchange effects
for such matrix elements should be very small.

By this discussion we finish the group-theoretical analysis of nuclear
matrix elements. In the next section we will discuss the results
of the numerical calculations of the various  observables of
six-nucleonic systems treating them as $\alpha$-2N systems.

\section{Six-nucleon system within the framework of the AMDM}

      In this section we discuss the AMDM application
to the six-nucleon systems ($^6$Li and $^6$He) treating them as the
$\alpha$-cluster  in  its ground
state with $L=S=T=0$ and two outer nucleons. The Jacoby coordinates
for this system together with the conjugated angular momenta are shown
in Fig.1.

As an input to the AMDM calculations, the three-body MDM functions
$\Psi(\vec\rho,\vec r)$ named as FUNCTION-92 in Ref.~\cite{fun92}
are used. These wave functions were calculated with a
large variational basis which includes all important (i.e. with the
weight P$\ge$0.1\%) components
with the partial angular momenta $l,\lambda \le 4$.

As has been stated in the Introduction,
not the intrinsic cluster wave functions but the corresponding
experimental data for the studied observables are used in the
calculations within the MDM. However, when
going from the MDM to the totally antisymmetric wave functions
one has to use explicitly the intrinsic wave functions of clusters
in analogy with the MCM.

    The major part of the calculations within the MCM
was performed using  the simple 0S$^4$ harmonic oscillator
shell model function for
the $\alpha$-particle. Of all possible nucleon correlations
only the breathing mode of the $\alpha$-particle or the $^3$H-p
($^3$He-n) channel was incorporated into
the most advanced calculations (see, for example,
Ref.~\cite{varga2,fuji,csoto}). For many static observables,
such a simplified version of the $\alpha$-particle wave function
does not affect the final result.  However, when large momentum
is transferred to a multicluster system, then the short-range
correlations inside the $\alpha$- particle become very important.

Unfortunately, it is known that the $\alpha$-particle wave function
calculated within a realistic dynamic approach does not reproduce
its Coulomb form factor at intermediate momentum transfer, when
evaluated in the impulse approximation.
To reproduce the experimental data either a very large contribution of
the isoscalar exchange current has to be assumed
or the existing wave function should be modified to reflect the more
complicated dynamic of this system.
It seems that such a large contribution of the MEC in the $\alpha$
-particle, and therefore in the core of the six-nucleon system,
is physically not adequate.
This is the reason why we decided to modify the $\alpha$-particle
wave function
and did it in a purely phenomenological way.
When constructing this wave function
we kept in mind that it should have the form
convenient for performing the calculation of matrix elements
in six-nucleon system with the antisymmetrized wave function of
eq.~(\ref{amdmk}).

Following these requirements a phenomenological $\alpha$-particle
wave function was constructed in Ref.~\cite{amdm}:
\begin{equation}
\Psi = \exp\left(-\frac{\alpha_1}{2}                     \label{fun_a}
\sum_{k=1}^{4}(\vec r_k - \vec R)^2\right) +
C \sum_{i=1}^{N}\exp\left(-\frac{\alpha_1}{2}
\sum_{k \neq i}(\vec r_k - \vec R)^2 - \frac{\alpha_2}{2}
(\vec r_i - \vec R)^2\right)\ ,
\end{equation}
with the parameters $\alpha_1=0.6144 fm^{-2}$, $\alpha_2=6.967 fm^{-2}$
and $C=-0.4506$.
Within the impulse approximation, this function fits the
$\alpha$-particle charge form factor in a  wide  region  of
momentum transfer.
This function was used in Ref.\cite{amdm}
to calculate the $^6$Li
form factors within the AMDM. A significant
improvement  of the
theoretical description of $^6$Li form factors at medium and high
momentum transfer was achieved by using this modified $\alpha$-particle
wave function.

The other advantage of the function (\ref{fun_a})
consists in its
similarity to the oscillator 0S$^4$ one. Their overlap
 ($<\Psi|\Psi_{osc}>$) is larger than 0.9.
Therefore one would expect that the qualitative
results of the group-theoretical analysis discussed above
will be valid in this case as well.

   So the model with antisymmetrization is completely defined and we are
ready to start  presenting the results of calculations within the model.
We will discuss below the following three versions of the model:
\begin{itemize}
\item the MDM,
\item the AMDM$_{\rm C}$ , where the normalization with a constant is
adopted, and
\item the AMDM$_{\rm K}$ ,  where  the  integral  kernel
$\hat K$ is used for the renormalization.
\end{itemize}

\subsection{Operators and matrix elements involved in  calculation
  of the observables of the six-nucleon systems.}

Three types of tensor operators are involved in our calculations
of the various observables of the six-nucleon  system. They are:

\begin{equation}
 O^{um_u ,tt_z} _{kw} (\hat r_j) = \tau_{tt_z} f(qr_j)
\sum \langle km_k wm_w :um_u \rangle
   \sigma_{km_k} (j) Y_{w m_w}(\hat r_j) ,  \label{32}
\end{equation}

\begin{equation}
O^{um_u ,tt_z} _{0w} (p_j) =   i\tau_{tt_z} f(qr_j)
\sum \langle 1m wm_w : um_u \rangle Y_{wm_w} (\hat r_j) p_{1m}(j)
  \label{33}
\end{equation}
and

\begin{equation}
     O^{um_u , tt_z} (\vec \sigma_j \vec p_j) =
iO^{um_u ,tt_z } _{0u}(\hat r_j) (\vec\sigma_j \vec p_j)  .
   \label{34}
\end{equation}
Here $\vec p$ is the nucleon momentum inside the nucleus and $q$
is the
momentum transfer to the nucleus;
$\tau_0$ and  $\sigma_0$ denote the unit operator $(k=0,  t=0)$
in the isospin and spin space, respectively,
and $\tau_{1z}$  and $\sigma_{1z}$ are the ordinary isospin and
spin operators $(k=1,  t=1)$, respectively;
$f(qr_j)$ is the observable scalar function associated with
the process which is being considered.

     We designate the nuclear reduced matrix elements
associated with the above given operators in the following
way:
\begin{equation}
<J_f T_f \| O^{um_u ,tt_z} _{kw} (\hat r_j) \| J_i T_i> =
[kwu]_{\Delta T} \ ,
\end{equation}
\begin{equation}
<J_f T_f \| O^{um_u ,tt_z} _{0w} (p_j) \| J_i T_i> =
[1wu;p]_{\Delta T}
\end{equation}
and

\begin{equation}
<J_f T_f \| O^{um_u , tt_z} (\vec \sigma_j \vec p_j) \|
J_i T_i> = [0wu;p]_{\Delta T} \ .
\end{equation}
They  are specified
according to the tensor structure of the corresponding operators.
To distinguish the nucleon momentum independent matrix elements
(\ref{32})
from the momentum dependent ones, (\ref{33}) and (\ref{34}),
the latter will be labeled
by an extra index p.
$\Delta T$ reflects the isospin selection rule in the nuclear transitions.
If the isospin of both the initial and
final states equals to zero then $\Delta T=0$, and if the isospin of one
state is T=0 and of the other is T=1 then $\Delta T =1$.
Only when both isospins are
equal to 1 we have both values;  $\Delta T =0$ and 1.

     In this paper we will discuss:
\begin{itemize}
\item
the $^6$Li and $^6$He observables in their ground state,
\item
the transition  $J^\pi T = 1^+0 \rightarrow 0^+1$ and vice versa
generated by various projectiles,
\item
the $\gamma$-transition to the $J^\pi T$ =$2^+$1 level and
\item
the transition $J^\pi T = 1^+ 0 \rightarrow J^\pi T = 3^+0$ .
\end{itemize}

Matrix elements
involved in the calculations of the corresponding observables
are the following:

     1. The ground state of $^6$Li ($\Delta T = 0$):

\begin{itemize}
\item
      i) charge and body (matter) rms radii -- $[000]_0$;
\item
     ii) quadrupole moment -- $[022]_0$;
\item
    iii) longitudinal form factor -- $[000]_0$ and $[022]_0$;
\item
     iv) magnetic moment -- $[101]_0$ and $[111;p]_0$;
\item
      v) magnetic elastic form factor -- $[101]_0$, $[121]_0$ and
$[111;p]_0$;
\item
     vi) the elastic pion scattering -- $[000]_0$, $[022]_0$, $[101]_0$
and $[121]_0$.
\end{itemize}

   2. The ground state of $^6$He ($\Delta T = 0$ and 1):
\begin{itemize}
\item
      i) rms charge radius -- $[000]_0$ and $[000]_1$;
\item
    ii) rms body radius -- $[000]_0$.
 \end{itemize}

 3. The transition from the ground state of $^6$Li to the $J^\pi T =0^+ 1$
level in $^6$Li or in $^6$He and vice versa:
\begin{itemize}
\item
      i) beta decay -- $[101]_1$;
\item
     ii) muon capture -- $[101]_1 , [121]_1 , [011;p]_1$ and $[111;p]_1$;
\item
    iii) form factor -- $[101]_1 , [121]_1$ and $[011;p]_1$;
\item
     iv) positive pion photoproduction -- $[101]_1$ and $[121]_1$.
\end{itemize}

   4. The gamma transition to the $J^\pi T = 2^+ 1$  level of $^6$Li:
\begin{itemize}
\item
      i) $[101]_1$ and $[111;p]_1$.
\end{itemize}

   5. The transition from the ground state of $^6$Li to the
$J^\pi T = 3^+ 0$ level in $^6$Li:
\begin{itemize}
\item
      i) gamma transition and electron scattering --
$[022]_0$, $[044]_0$ and $[112;p]_0$;

\item
  iii) pion scattering -- $[022]_0$, $[044]_0$, $[12u]_0$ and $[14u]_0$.
\end{itemize}

Among the above listed matrix elements only  $[000]_0$, $[022]_0$,
$[044]_0$ and $[1wu;p]_0$  are of pure
spatial type. For overlapping clusters, as has been shown in
Ref.\cite{amdm}, ordinarily
the magnitude of the exchange effects  for a pure spatial operator
is associated with the quantity $\delta=1-Q^2$, where $Q^2$ is given by
eq.~(\ref{Q}). The
MDM functions \cite{fun92}, which are used
throughout this paper, were constructed in a way to minimize the admixture
of components with Pauli-forbidden Young-schemes [f] = [51] and [6] (see
Ref.\cite{amdm}). For this reason the values of $Q^2$ are rather
close to 1 and
therefore the exchange effects are usually not large. That is why such
observables
as $^6$Li charge and body radius, quadrupole moment and $^6$He body radius
should have very close values in all versions of the model.
The direct calculations have demonstrated (see Table 1) that the
renormalization
effects in this group of matrix elements is indeed small.

  Matrix elements $[101]_0$ and $[101]_1$, in the long-wave approximation
where $f(qr_j)=Const$, are of pure spin and spin-isospin types,
respectively.
The kernel renormalization restores exactly their values, altered by the
C-renormalization. At larger momentum transfer, as follows from Table 1,
they depend on the renormalization procedure.
Now, let us turn to the discussion of the renormalization effects on
various observables of the six-nucleon nuclei.

\subsection{The static properties of the six-nucleon systems}

In Table 2 the results of calculation of the static properties of the
$^6$Li (columns 2--4) and $^6$He (columns 5 and 6) are presented.
\subsubsection{Ground state of $^6$Li}

Only one matrix element contributes to the rms radius and the quadrupole
moment of $^6$Li. Sensitivity of these observables to the version of the
renormalization procedure have been discussed above.
Here one can only add that the calculated rms radius is in agreement with
the experimental data
whereas the quadrupole moment is far from the measured one.
The theoretical prediction for the value of the quadrupole moment is
a special problem and it is beyond the scope of this paper.
It seems that one should take into account the D-component of the
$\alpha$-particle wave function to come close to the experimental value
of quadrupole moment of $^6$Li.
It is worth mentioning that
in all large-scale calculations of this moment
(see, for example, the discussion
in Ref.\cite{amdm} and references therein)
its predicted value is very close to that
obtained in this paper.

One can see from Table 2 that both the exchange and renormalization
effects for all static observables  of six-nucleon system
are not large.

\subsubsection{Gamma transition to the $J^\pi T=2^+1$ level in $^6$Li}

     The exchange effects in the static characteristics can increase
if the leading matrix element of  the  corresponding  operator  is
suppressed due to nuclear structure effects. Such a  situation
occurs in $^6$Li for the M1  $\gamma$-transition  from  the  ground state
to  the
$J^\pi T=2^+1$ level. Here the magnetization current
matrix element $[101]_1$ is strongly suppressed due to the structure of
the corresponding wave functions (the corresponding operator $O$ is unable
to change the total angular momentum L by two units) and the main
contribution
comes from the  convective  current, $[111;p]$.  So,  this  transition
belongs to the scissor mode. The prediction of the
MDM and of two
versions of the AMDM differ noticeably from  each  other:

      $\Gamma_{\gamma_0}$  (eV) = 0.29 (MDM),
                             0.38 (AMDM$_{\rm C}$) and
                             0.34 (AMDM$_{\rm K}$).
The experimental value is  ($0.27 \pm 0.05$) eV.

The result obtained within the framework of the MDM is closer to that
of the AMDM$_{\rm K}$, than to the AMDM$_{\rm C}$  one.
The AMDM$_{\rm K}$ result deviates a little from the measured value.
But it is expected that in this hindered transition, the mesonic exchange
currents could play a very important role. So it will be interesting to
continue to investigate this problem further.
Notice that for many other
observables the AMDM$_{\rm K}$ results are also close to those of
the MDM. That is why in  these
cases the MDM has been very successful in describing these observables.

\subsubsection{Ground state of $^6$He}

  The neutron halo nucleus $^6$He is one of the interesting peculiarities
of  the  six  nucleon system. The outer neutron pair, forming
this halo, extends to large distances and
is localized in the so called quasiasymptotical region.
The charge distribution in $^6$He is associated with two types of matrix
elements -- $[000]_0$ and $[000]_1$, contrary to the matter distribution
which is associated only with the isoscalar part.  So, we can compare
the results obtained for
two similar operators in the coordinate space which, however,
have different isospin structure.
This comparison is done in Table 2.

When  the kernel  renormalization  is adopted within the AMDM,
the rms radius of $^6$He is changed by less than  1\% when  compared
with the
MDM case. Let's define, following Ref.\cite{csoto}, the size of the halo
in $^6$He as
the difference between the proton and neutron
radii $r_{halo}=<r^2_n>^{1/2} - <r^2_p>^{1/2}$.
It  changes by
about 7\% when one goes from the MDM to the  AMDM$_{\rm K}$  despite
its quasiasymtotic nature.
The exchange and renormalization effects for the halo
in $^6$He are not well pronounced. As a rule one  should  keep  in
mind this effect when calculating the radius of the halo
in heavier nuclei, where the antisymmetrization between nucleons located
in different clusters is neglected.   More  complicated
intrinsic structure of the constituent cluster in such systems, $^9$Li
cluster in $^{11}$Li as an example,
would increase the exchange effects and, as a result, would
increase the calculated halo radius.

   The antisymmetrization procedure with renormalization to a
constant distorts the asymptotic behavior of the wave function.
As a result a nonphysical exchange effect
at large distances arises in this case. This is the reason  why
the body rms radius
of $^6He$ changes when going from the MDM to the AMDM$_{\rm C}$.
To avoid this unphysical effect one should deal with the kernel
renormalization from the beginning.

\subsubsection{Beta decay of $^6$He to $^6$Li}

The $^6$He $\beta$-decay rate is determined by the spin-isospin
Gamow-Teller
operator $\hat\sigma\hat\tau$:
\begin{equation}
ft = \frac{2 ft_{0^+\rightarrow 0^+}}{(g_A/g_V)^2 |M|^2} .  \label{beta}
\end{equation}
We use the following values \cite{weak_cons}
for the constants in eq.~(\ref{beta}):
$2 ft_{0^+\rightarrow 0^+} = 6144\ s\mbox{ and } (g_A/g_V)=-1.259$.
The results of the $ft$ calculation are the following:
              $ft$ = 796 s (MDM and AMDM$_{\rm K}$) and
                     772 s (AMDM$_{\rm C}$).
The experimental value of $ft$ is 813 s. Keeping in mind that
there should be some meson exchange current contribution,
it seems that the calculated value of
the $\beta$-decay matrix element is somewhat overestimated.
The AMDM$_{\rm K}$ result is exactly equals to the MDM one,
as it should be.

\subsection{Electromagnetic form factors}

\subsubsection{Longitudinal form factors}

     The results of the calculations of the longitudinal elastic form
factor of $^6$Li are given in Fig.2. Starting from this
figure and in all subsequent ones we present the results
in the following way. By the solid line we show the results obtained
within the framework of the AMDM$_{\rm K}$, by the dashed line --- within
the AMDM$_{\rm C}$ and by the dotted line --- within the MDM.
The full elastic form factor consists of the sum of the contributions of
the C0 and C2 terms. However the C2 form factor is
visible only in the region of the C0 minima which is located at
about 3 $fm^{-1}$. That is why we are not showing them separately.
All longitudinal form factors for transitions between nuclear
levels  with  isospin $T=0$ are  associated  with pure   spatial
operators.
At low momentum transfer the elastic charge form factor of $^6$Li is
determined by its
rms charge radius which is practically not
affected by the nucleon exchange. More importantly, the
exchange effects remain very small up to a high momentum transfer.
The effects of the kernel renormalization are too small to be seen
in the Fig.2. So, one can safely use the MDM for this elastic
longitudinal form factor calculations.

As it had been discussed earlier in Ref.~\cite{amdm} and confirmed by
the present calculation, in order
to reproduce the $^6$Li elastic form factor it is  very
important to use the correlated $\alpha$-cluster wave function
which reproduces the $\alpha$ form factor. Otherwise, the calculated high
momentum part of the $^6$Li elastic form factor will be much lower
than the measured one.

     The antisymmetrization results in the
mixing of
the components of the three-body functions which have  different
partial angular momenta $l$ and $\lambda$ but the same  total $L$.
In the $^6$Li ground
state wave function this situation is not well pronounced, because the
dominant weight is associated with the $l=\lambda=L=0$ component.
In  the
wave function of the $J^\pi T = 3^+0$ level in $^6$Li there are already
two such components:
$l=2,  \lambda=0$ and
$l= 0,  \lambda = 2$ with the total angular momentum $L=2$ and spin $S=1$.
The antisymmetrization with renormalization results in their additional
mixing, which
differs essentially from the MDM mixing. For the same reason
there is noticeable difference in the wave functions of this level
in the two versions of the AMDM - with and without kernel renormalization.
Thus the modification of this inelastic longitudinal form factor
due to the antisymmetrization and renormalization should be rather large.

Fig.3 displays the  $\sqrt{B(q^2,E2)}$---the
square root of the reduced quadrupole transition strength to this
level at
low momentum transfer region. Just this multipole dominates in
the form factor for this transition. The kernel renormalization effects
are visible already at zero
momentum transfer due to this additional mixing.
At photon point $B(E2,1^+0\rightarrow 3^+0)$ in $e^2fm^4$ equals
to  20.05, 15.25 and  20.71 for the
MDM, AMDM$_{\rm C}$ and AMDM$_{\rm K}$ models, respectively.
The experimental
value is $21.8 \pm 4.8$ \cite{ajzenberg}. So, more precise measurement
of this observable seems to be needed to test the theoretical
models.
     Fig.4 displays the form factor for this transition in a
broad region of momentum transfer. In logarithmic scale, it is not easy
to see the difference between the three versions of the model.
In the region of the first maximum this difference is about the same as
in the region of very low
momentum transfer. The magnitude of the calculated form factor is
lower than the experimental one. The reason for this is not clear enough.
At high momentum transfer, for example, it is possible that
noncentral components of the $\alpha$-particle  wave  function, which
are not taken into account at all, start to play a role.
So, this discrepancy should be a subject
of further studies.

\subsubsection{Magnetic form factors}

     The earlier MDM calculations of the $^6$Li magnetic form factors
in Ref.~\cite{23.} have demonstrated
quite
clearly that despite of a good agreement with experimental data in
the region of the first maximum, the theoretical  result  for  the
region of the second maximum lies much below the experimental  data
for any choice of NN and $\alpha$-N interaction \cite{23.,24.}.
At the same time, the studies carried out
in  Ref.~\cite{amdm}
showed qualitatively  that  the  antisymmetrization  is  of  great
importance when one calculates the $^6$Li transversal form factors.
Both the elastic and  the  inelastic transversal form factors
to the $J^\pi T=0^+1$ level
are governed by the  M1  multipole  only.
The antisymmetrization of the wave function leads to a strong
increase of the form  factor  within  the  region  of  the  second
maximum, thereby improving the agreement  with  the experimental  data
substantially. This is demonstrated in Fig.5  and Fig.6.

     At low momentum transfer the M1  operator  is proportional
either to the operator of the magnetic moment (in the case of  the
elastic  scattering)  or  to  the  GT-operator of $\beta$-decay
and  has mainly a spin-isospin component.
Here we use the fact that the convective current contribution is small.
That is  why  the  exchange  effects  are  very
small in this region of momentum transfer. However, already at the
second maximum their  role  becomes  crucial
because the spatial part of the M1 operator starts to play its
decisive role.

     It is necessary to stress that in the AMDM$_{\rm K}$  version
the position of the minimum is reproduced  very precisely.
The other interesting effect of kernel renormalization
consists in stabilization of the AMDM results for
intermediate momentum transfer ($q\le 2.5 - 3.0 fm^{-1}$):
the magnitude of the form factors
in the region  of  the
second maximum becomes almost independent on the  details  of  the
$\alpha$-cluster wave function. It means that in this region the
exchange and short-range correlation effects are well separated
from each other. This result  is  very  important  because  our
$\alpha$-cluster wave
function  has a pure  phenomenological origin.

    At larger momentum transfer an interference between two types
of nucleon correlations (exchange and short-range)  starts to play
its decisive role. The consistent treatment
of both effects results in appearance of a third
maximum in the M1 form factors at about $q^2 \simeq 20 fm^{-2}$.
Again as it happened with the second maximum, the AMDM$_{\rm K}$ version
provides more stability of the third maximum in the magnetic form factor
as compared to the AMDM$_{\rm C}$ version.

Elastic M1 form factor has an isoscalar nature. Therefore,
the exchange currents in this case are not very strong and should
not alter considerably the results for small and moderate momentum
transfer (see discussion in Ref.\cite{amdm}).
The inelastic M1 form factor has an isovector origin and it
is  somewhat
lower  than the experimental data, already at lower momentum transfer.
One can assume that this is a real indication for the mesonic exchange
currents in this transition. At higher momentum transfer their
contribution should increase.

It seems to us that it is very promising to continue the experimental
study of $^6$Li form factors and to cover the high momentum transfer
region. This can be an interesting topic, for example, in CEBAF.

\subsection{Gamow-Teller transition in muon capture}

The Gamow-Teller transitions in muon capture on nuclei are interesting
from the following points of view:

- how well the nuclear models are able to predict their strength at
intermediate momentum transfer,

-what is the magnitude of the induced pseudoscalar coupling $g_P$
in nuclei,

-how large are the mesonic exchange currents  at this momentum transfer.

Up to now these problems have been analyzed in great details only in the
few-body systems. $^6$Li gives the other opportunity to study the above
discussed problems in a nucleus with twice as many nucleons as $^3$He. 
Here, all these effects can be enhanced.

The momentum transfer in muon capture on $^6$Li with formation
of $^6$He in its ground state is equal to $|\vec q|$   =
$E_\nu/c$  = 100.7 MeV/c. This value is in the region of the first
maximum in the inelastic electron scattering
form factor (see Fig.6) to the J$^\pi$ T = 0$^+$ 1 level of $^6$Li.
Four matrix elements form two transition amplitudes, $T_1$ and
$T_2$, to the ground state of $^6$He:

\begin{equation}
T_1 = {2G_A [101] \over 3}\{1 - \Delta T_1\}      \label{T1}
\end{equation}
and

\begin{equation}
T_2 = (2/9)^{1/2} (G_A - G_P) [101] \{1 + \Delta T_2 \}.    \label{T2}
\end{equation}
Here we have isolated the dominant matrix element [101] from the rest.
$\Delta T_1$ and $\Delta T_2$ are built up from the three others,
[121], [111p] and [011p]:

\begin{equation}
  \Delta T_1 =  \sqrt{1/2}  {[121] \over [101]} +
\sqrt{3/2} (g_V / G_A )  {[111p] \over {M_N [101]}}     \label{DT1}
\end{equation}
and

\begin{equation}
  \Delta T_2 =  \sqrt{2}  {[121] \over  [101]} +
{{3g_A} \over {G_A - G_P}} {[011p] \over {M_N [101]}} .   \label{DT2}
\end{equation}
$G_A$ and $G_P$ are some combinations of the weak interaction
coupling constants. They are given in Appendix A together
with the expression for the capture rate.

Two observables are of experimental interest in this partial muon
capture transition --
the capture rate $\Lambda$, when the hyperfine states are populated
according to their statistical weights,
and
the ratio of the capture rates from the upper, $\Lambda_+$ ,
and the lower, $\Lambda_-$ , hyperfine states of the $^6$Li mesic atom.

The capture rate is proportional to the sum of $T_1^2$ and $T_2^2$,
the ratio $\Lambda_+/\Lambda_-$ is given by the following expression:

\begin{equation}
\Lambda_+/\Lambda_-=\frac{G_P^2}{\{3 G_A-G_P\}^2}
\left\{1+\Delta_2\right\},
\end{equation}
where

\begin{equation}
\Delta_2 =\frac{3(G_P-G_A)}{2 G_P}\left\{1-\frac{3}{2X}\left[1-
\frac{G_P}{3G_A}\right]\right\}^{-1}           \label{delta2}
\end{equation}
and

\begin{equation}
X=\frac{\Delta T_1 +\Delta T_2}{1+\Delta T_2}. \label{X}
\end{equation}
It is through $X$ that the nuclear structure is manifested in
the ratio of the capture rates $\Lambda_+/\Lambda_-$.
In the limit $\Delta T_1$ and $\Delta T_2$ $\rightarrow$ 0,
$X$ and $\Delta_2$ are equal to zero and this ratio
is given by the coupling constants only showing a strong dependence on
$g_P$. That is why this ratio is a subject of detailed analysis
in many nuclei (see Ref.~\cite{ratio}).

$\Delta T_1$ is independent of $g_P$. After fixing the weak interaction
constants according to Appendix A one arrives to the values of
$T_1$ and $\Delta T_1$ listed in Table 3. As follows from Table 3, the
combination of matrix elements which forms $\Delta T_1$ is
sensitive to the nuclear model. But again, the result in
the AMDM$_{\rm K}$ version is closer to the MDM one. The difference
is about
10\%. The full amplitude $T_1$ is less sensitive to the model because
$\Delta T_1$ itself is about half of a per cent. The combination of
the matrix elements which forms $\Delta T_2$ is not very
sensitive to the model as well. But its relative weight in the
full amplitude $T_2$ is about 15 \%.

In both cases it is important to take into account the velocity
dependent matrix elements (the corresponding operator is proportional to
the momentum of nucleon inside the nucleus) . If one neglects
them (see last column of Table 3), as it is done in some papers, then
both terms, $\Delta T_1$ and $\Delta T_2$, will be changed drastically.

The ratio of the hyperfine capture rates depends on the nuclear
model through $X$ 
sensitivity of the amplitudes and their combinations
to the nuclear model we give
in Table 3 their magnitudes for fixed value of the induced pseudoscalar
coupling constant ($g_P/g_A$ = 7). Finally, for the same value of the
coupling constant we give the capture rate and the ratio
$\Lambda_+/\Lambda_-$.

Fig.7 demonstrates the calculated capture rate as a function of
$g_P$/$g_A$ together with experimental data \cite{muon6li}.
They are far from each other. Even if one increases arbitrarily the
theoretical capture rate by about 10\% , to take into account
in this way the meson exchange current contribution,
the situation improves only slightly.
But nevertheless, the value of the induced pseudoscalar constant
which is needed  to reproduce the experimental data is lower than
that which  follows from the Goldberger-Treiman relation.
It seems very desirable to remeasure this capture rate with a better
accuracy to insure that situation is not due to the existing
experimental data.

Fig.8 demonstrates the predicted values for the ratio of the capture
rates as a function of $g_P/g_A$ for the three versions of
the model.
The absolute value of this ratio appears to be the same
within about 3\%. It means that the uncertainty
from the nuclear structure side is minor. As the next step, one should
incorporate into the calculations the meson exchange currents, which
can modify the result obtained for the ratio of the capture rates.

\subsection{Partial transitions in pion scattering and pion
                  photoproduction off $^6$Li}

     After successfully employing the developed model for
description of the $^6$Li traditional observables
let us turn our attention to the:
\begin{itemize}
\item  pion  elastic  and  inelastic  scattering  with
the excitation of the $J^\pi T= 3^+0$ level,
\item pion photoproduction with the formation of $^6$He in  its
ground state ($J^\pi T=0^+1$).
\end{itemize}
These reactions are selected for the following reasons. Pion scattering
on polarized target has already been carried out and the vector
analyzing power i$T_{11}$ for the ground state and the $J^\pi T = 3^+0$
level
transitions was measured.
Nuclear models can be tested  in  a more  thorough  way
via the polarization observables
than via the differential  cross  sections,  provided  that  the
reaction mechanisms are properly taken
into account. So,  pion  scattering  can  be  considered  as
complementary  to electron  scattering in probing
details of the $^6$Li structure. One can invert the formulation
of the problem --- if the nuclear structure is well determined,
then the other ingredients of the theory can be checked.

     As to  the  pion partial photoproduction off nuclei it
is expected that
very soon the data with polarized photons will be
available as well. It will be  data on the beam asymmetry:
\begin{equation}
             \Sigma (\theta) =
\frac{\sigma ^\perp (\theta)  -
\sigma ^\| (\theta)}{\sigma ^\perp (\theta)  +
\sigma  ^\| (\theta)}  ,
\end{equation}
where $\theta$  is the polar  angle  of  the  outgoing  pion,
$\sigma ^\|$
is  the
differential cross section for the photon polarization along the $x$
axes, and $\sigma ^\perp$  --- along the $y$ axes.
The pion  momentum $\vec q$
is  directed
along the $x$ axes, the photon momentum $\vec k$ along the $z$  axes
and  the vector
$[\vec k \times \vec q]$ along the $y$ axes.
We will demonstrate that, depending on the momentum
transfer (pion emission angle), there are regions sensitive either
to the nuclear model or to the details of the reaction mechanism.
At present there are polarization measurements in pion photoproduction
when nucleon knocked out from nucleus
.
     Pion  scattering  and pion photoproduction on light nuclei were
described   very
successfully within the framework of DWIA in  momentum  space  (see
for example Ref.~\cite{MK,E88}). However, in some cases
(including elastic and inelastic to the $3^+0$ level
scattering  in  $^6$Li)  one  has  to  use  the
coupled channel method (see for example Ref.~\cite{R94}).

     In this paper we will continue to exploit the coupled channel
method for pion scattering to the ground state,
the $J^\pi T$ = $1^+0$ and the $3^+0$
levels,  in  line
with Ref.~\cite{R94}. Pion photoproduction will be analyzed within the
framework of the DWIA, in line with Ref.\cite{E88}.
Some basic formulas for both processes are given in Appendix B.

Among all low lying states of $^6$Li the elastic and the inelastic
to the $3^+0$ cross sections are measured  for
several energies. Cross sections to these levels are large due to the
isoscalar nature of the nuclear transitions. More or less complete
data, which include
the polarization observables, exist for  a  few  energies  of  the
incoming pions. One of such energy is $T_\pi$ = 134 MeV. That  is
why  we
will concentrate all our attention mainly on the data at this energy.

\subsubsection{Elastic and inelastic to the $J^\pi =3^+0$ level pion
scattering}

     Elastic and inelastic to the $3^+0$ level pion scattering off
$^6$Li are
governed by the spin independent part of the elementary $\pi$N
scattering amplitude. The spin-dependent part of pion-nucleon
amplitude is much smaller and does not show up in the differential
cross section. In this section we compare our results with the
experimental data of Ref.\cite{R94}.

Fig.9  displays  the elastic cross  section  for  134  MeV
pions. All three versions of the model  give
similar results. The situation is the same as in the case of electron
scattering where the calculated longitudinal form factors appeared
very close to each other.
The point  is  that
in pion scattering  at this energy the  maximal  momentum  transfer
does  not
exceed 2.5 $fm^{-1}$ where these three versions of the model predict
very close values for the matrix element --- see Table 1.

     In the case of the $3^+0$ level the AMDM$_{\rm C}$ version  at
low  momentum transfer region (small angles) gives the  lowest  result
just as
in the electron scattering case (see Fig.10).  The
MDM and AMDM$_{\rm K}$ versions predict a higher value which
is more close to the experimental data.
Predictions of all three versions of the model
in the region of the minimum differ from each other significantly.

     In both cases one can say that agreement between theory  and
experiment  is  quite  good  and  some  deviation  of  the theory from
experiment can  be  due  to  some  shortcomings  associated with  the
reaction part. For example in this calculation the pion absorption is
ignored. Of course this process can not  play  a  decisive
role, but it can have some effects.

     Contrary to the differential cross section the vector polarization
$T_{11}$ is caused by interference between the spin-dependent and
the spin-independent parts of the pion-nucleon amplitude.
     To have some feeling of  $T_{11}$
it is  instructive
to use the plane wave approximation, following Ref.\cite{Ritt}.
In the elastic longitudinal electron scattering
form factor  one could
safely neglect the $L$=2 matrix elements. The same is true for the pion
scattering. Then in the plane wave approximation one gets

\begin{equation}
iT_{11}(d\sigma / d\Omega)
\sim [000][101]Im[f_{\pi N}(s=0)f_{\pi N}^\ast (s=1)].
\end{equation}
Here $f_{\pi N}(s=0)$ and $f_{\pi N}(s=1)$ are the spin-independent and
spin-dependent  parts  of
the elementary $\pi N$ scattering amplitudes, respectively.
From this expression it follows that $iT_{11}$ changes its sign where
the M1 form factor does. The largest value of the vector analyzing power
corresponds to the minimum in the differential cross section of
pion scattering.
The $Im [f_{\pi N}(s=0)f_{\pi N}^\ast  (s=1)]$
changes its sign at $\theta$ larger than 120$^o$
for $T_\pi >$ 120 MeV. Thus the vector analyzing power
is a complicated function of the scattering angle.

The vector analyzing power $iT_{11}$ for the transition to the $3^+0$
level in the plane wave approximation  looks  very  similar:

   \begin{equation}
iT_{11}(d\sigma / d\Omega) \sim
[022] \{[122] + (32/49)^{1/2} [123]\} Im[f_{\pi N}(s=0)
f_{\pi N} ^\ast (s=1)].
  \end{equation}

     The results of comparison between experimental data and the three
versions of the model is given in Fig.11 for the elastic scattering and
in Fig.12 for the transition to $J^\pi T= 3^+0$ level.

        Contrary to the differential cross section the vector analyzing
power is more sensitive to the nuclear model. All three versions of
the model used in the calculations give different results.
The AMDM$_{\rm K}$ version
explains the shape of the vector polarization as a function of
scattering angle. There is some discrepancy in the magnitude of the
vector analyzing power. This discrepancy can be caused by the approximate
treatment of the pion-nuclear dynamic.
But what emerges from the calculations
is that the nuclear model is already reliable enough so one can begin
to improve upon the
dynamic. Simultaneously it seems to be very useful to have more precise
experimental data on this quantity.

\subsection{Pion
 photoproduction to the $^6$He ground state.}

Pion photoproduction at threshold is dominated by the Kroll-Ruderman
spin-flip term. So in the pion photoproduction on $^6$Li with formation
of $^6$He in its ground state the same matrix elements governing the
transition as in muon capture and in electron scattering. So at threshold
one expects a successful description of experimental data.
In Fig.13 the calculated results within the framework of
the AMDM$_{\rm K}$ are compared to experimental data of Ref.\cite{P1,P2}
at 200 MeV energy of the incoming photons. At this energy the momentum
transfer is covered from 0.5 $fm^{-1}$ at $25^o$ to 1.5 $fm^{-1}$ at
$135^o$. At this end point pion photoproduction is governed by the
form factor of electron scattering in its minima, where it is very
difficult to expect a precise theoretical description of the
momentum dependence. That
is why at backward angles there is a serious disagreement between
theory and experiment. As to the forward angles one sees a nice agreement
with experimental data.

At higher energies of incoming photons,
say at 320 MeV, contribution of $\Delta$-isobar becomes very important
and it brings the spin-independent part into the amplitude.
At this energy the difference in the differential cross section
between the three versions of the model is very small at
forward angles which correspond to low momentum transfer. Fig.14
demonstrates this in a clear way. Predictions become different from each
other starting from the angle $\theta$ at about $60^o$.

The beam asymmetry at forward angles is also
weekly sensitive to the model --- see Fig.15.
At these angles a strong sensitivity of the beam asymmetry
to the $\Delta$-isobar property inside a nucleus
is found. In Fig.16 the beam asymmetry is plotted as a function of the
$\Delta$-isobar mass, $M_\Delta$, in nuclei. A 5\% deviation of this mass
from its free value changes the beam asymmetry noticeably. This happens
in the region of angles from 20 to 40$^o$ where the cross section is not
small. So, this result can be considered as a motivation for measuring
the beam asymmetry in this transition.
Variation of the $\Delta$-isobar mass
affects the differential cross section mainly at backward angles
(Fig.17). However at these angles there is a strong sensitivity to the
nuclear model as well.

\section{Conclusion}

A modified version of the Multicluster Dynamic Model was proposed
to construct completely antisymmetrized wave functions of
multicluster systems. In the model the intrinsic
wave functions of the constituent clusters were inserted explicitly,
and as a result the model
became closer to the MCM, where the motion of all nucleons is taken
into account on equal ground. However, contrary to the MCM,
the cluster-cluster potentials used are those constructed on the basis of
phase shifts analysis of the free cluster
scattering and thus the on-shell properties of these potentials are
taken into account in the most precise way. Such a modification,
nevertheless,
retains the easy application of the MDM and at the same time allows
us to incorporate
into the calculations the realistic wave functions which describe the
motion of the nucleons inside the clusters.

A group-theoretical analysis of matrix elements for arbitrary multicluster
systems was developed. On its basis all operators of  single-particle
origin have been classified according to their response to the
antisymmetrization
procedure. A group of operators is found whose matrix elements are
strictly conserved when going from the MDM to the AMDM$_{\rm K}$,
but not to
the AMDM$_{\rm C}$. This analysis allows to predict, before the
numerical calculations are carried out,
how essential will the exchange effects be for the particular
nuclear characteristic. Simultaneously, it was shown that the kernel
renormalization removes the spurious effects which had appeared when
the renormalization to constant was adopted.

Detailed numerical calculations of various observables
in the six nucleon systems were performed.
These nuclei are particularly singled out in nuclear physics because
they already exhibit quite properly the typical properties of all
light nuclei, but at the same time the number of nucleons is
not too large
to prevent an accurate and consistent calculations starting from
the realistic interactions.

Within the framework
of the AMDM$_{\rm K}$
a better description of many six-nucleon observables has been achieved.
The electron scattering form factors are those obtained with
high precision.
One of the interesting effects of the kernel renormalization
consists in the stabilization of the AMDM results for
intermediate momentum transfer ($q\le 2.5 - 3.0 fm^{-1}$):
the magnitude of the magnetic form factors
in the region  of  the
second maximum became almost independent on the  details  of  the
$\alpha$-cluster wave function. It means that in this region the
exchange and short-range correlation effects are well separated
from each other. This result
is  very  important for the accuracy
of the model because  our $\alpha$-cluster wave
function  has a pure  phenomenological origin.

    At larger momentum transfer an interference
between two types of nucleon correlations
(exchange and short-range) starts to play its decisive role.
The consistent treatment
of both effects results in the appearance of a third
maximum in the M1 form factors at  $q^2 \simeq 20 fm^{-2}$.
And again the AMDM$_K$ version provides some stabilization of
this maximum.

The model developed was applied to muon capture, pion scattering and
pion photoproduction. It succeeded in providing a good description of
the differential cross sections in pion scattering. What is more
important, it allowed us to
reproduce at least the shape of the vector polarization which is more
sensitive to the details of nuclear structure. Thus we have demonstrated
that the AMDM$_K$ allows us to accurately describe tiny details of
the nuclear structure which, nevertheless, are essential for
the interpretation
of the polarization phenomena in nuclear reactions.

In the case of
pion photoproduction, a region of angles was found where the beam
asymmetry depends only slightly on the nuclear model and very
strongly on the mass of the $\Delta$-isobar inside nuclei. Experimental
data in this region of angles, if they appear, can provide fresh insight
into this problem.

The calculations performed have demonstrated that the results of the
group-theoretical
analysis are valid in a more realistic case as well,
when the cluster wave functions are more complicated as compared to
the SU(4) and SU(3) scalars.

There is one more aspect of this paper which can be formulated
in the following statement: after introducing the kernel renormalization
the model became more universal and can be applied now to more heavy
multicluster systems, contrary to the MDM.

The group-theoretical analysis carried out in this paper  offers
promise as
a processing tool and can be used in many other situations.
Here we would like to demonstrate two more areas of its application.
For a certain (and rather wide) type of light
systems the many nucleon wave functions constructed within the framework
of the oscillator shell model can
be rewritten in  bi- \cite{WK} or multicluster \cite{TC3} representation.
In other words such a wave function can be represented by eq.(\ref{amdmk})
where the $\alpha$-particle, $^{16}O$, etc., and nucleons act as clusters.
So the group theoretical analysis
may be applied to such systems as well. From this analysis one
is able to get
early insight on the role of the nucleon exchange in the discussed
systems and to estimate its effect arising from the
antisymmetrization between nucleons in different clusters.
We will not discuss all these issues because
they are beyond the scope of this paper.

There is another interesting field of application for the developed
scheme of the group-theoretical analysis.
The case in point is the problem known as quarks inside the nuclei.
If one starts from the three-quark representation of the individual
nucleons in nuclei (and does not take into account contribution of "sea"),
then the question whether one should take into account the quark
degrees of freedom of nuclei, or it would be enough to consider only
the nucleons as constituents of nuclei, depends substantially on the
magnitude of the quark exchange effects.

Indeed, as we have seen, in the multicluster system any process
without the breaking down of the constituent clusters can be
described within the
framework of the approaches where the clusters themselves are treated as
an elementary particle. Such an approach is valid when the
exchange effects between clusters are small.

To estimate the quark exchange effects in nucleus it is
again convenient to use the methods of group-theoretical analysis.
Let us discuss very qualitatively two conclusions which follow from
such an analysis.

1. The exchange effects should be larger for observables which are
described
by the full tensor operator. However even in this case the quark
exchange effects in nuclei are usually rather small because of
high internucleon distances. That is why the quark
degrees of freedom are seen in such processes only at very high
momentum transfer.

2. Simultaneously from our analysis (by reduction ab absurdum)
it follows that the quark effects in nuclei
should be more pronounced in processes where the break down
of the quark clusters or their excitation take place. Such processes are
enhanced due to the exchange effects \cite{TC1}.
One interesting example is already considered in Ref.\cite{TC2},
where the nucleon resonance as a spectator participates in the reaction.
Though the weight of the components with the nucleon resonance
in the  nuclear ground state wave function is very small,
observation and investigation of the
corresponding reaction is important due to
more transparent manifestation of the quark degrees of freedom
in nuclei in this case.

Thus, we have shown in this paper that the group-theoretical
analysis can open a new way for general investigation
of exchange effects in various
nuclear  and subnuclear processes. Some specific examples have been
discussed, for example, in  Ref.\cite{TC2}.

\acknowledgments

This work is supported in part by the National Science Foundation
under Grant \# PHY-9413872 and Russian Foundation of
Basic Research Grants \# 96-02-18691.

R.A.Eramzhyan would like to express his thanks to the Cyclotron
Institute at Texas A$\&$M University for the kind hospitality.

The authors would like to thank A.A.Chumbalov and S.S.Kamalov for
cooperation
in the initial stage of this work concerning the pion physics,
S.Shlomo for careful reading our manuscript and many valuable
suggestions and
V.I.Kukulin and V.N.Tolstoy for the numerous fruitful discussions.

\appendix

\section{GENERAL EXPRESSIONS FOR MUON CAPTURE RATES}

  The muon capture rate is expressed via the amplitudes given by
eqs.(\ref{T1}) and (\ref{DT2}) in the following way:

\begin{equation}
\Lambda  = 8\pi^3 \alpha^3 (m_\mu /m_e )^5 ln2 \eta (E_\nu) Z^3 R_Z
{1 \over {2ft_{0^+ \rightarrow 0^+}}}
 {{(2J_f + 1) \over
(2J_i + 1)} \Bigl\{T^2 _1 + T_2 ^2 \Bigr\}} .
\end{equation}
Here
\begin{equation}
\eta (E_\nu ) = (E_\nu /m_\mu )^2
\left\{1 - {E_\nu \over {m_\mu + M_i}}\right\}
\left[{1 \over {1 + m_\mu / M_i}} \right]^3  .
\end{equation}
$R_Z$ = 0.92
is the dimensionless quantity and is equal to the ratio between
the averaged muon wave function squared over the nuclear volume
and its value for the point nucleus at the origin,
Z is the charge of $^6$Li and $M_i$ is its mass. $M_N$ is
the nucleon mass.

After eliminating the main matrix element [101]
from the amplitudes $T_1$ and $T_2$ one comes
to the following expression for the sum of $T_1^2$ and $T_2^2$:

\begin{equation}
T_1 ^2 + T_2 ^2  = {2G^2_{G-T} [101]^2 \over 3} \{1+\Delta_1\}
\end{equation}
where
\begin{equation}
   G^2 _{G-T} = G^2 _A + G_P ^2 /3 - {2G_A G_P}/3
\end{equation}
is the square of the Gamow-Teller constant in muon capture,

\begin{equation}
\Delta_1=\frac{2 G^2_A}{3 G^2_{G-T}}
\left\{\Delta T^2_1-2 \Delta T_1\right\}
+ \frac{(G_A-G_P)^2}{3 G^2_{G-T}}
\left\{\Delta T^2_2+2\Delta T_2\right\}
\end{equation}
and

\begin{equation}
G_A(q^2) = {g_A (q^2) - 4.706 g_V (q^2)}{E_\nu \over {2M_N}} ,
\end{equation}

\begin{equation}
G_P(q^2) =\{ g_P (q^2) - g_A (q^2) - 4.706 g_V (q^2)\}
{E\nu \over {2M_N}},
\end{equation}
where $g_A$ is the axial-vector coupling constant,

\begin{equation}
  {g_A (q^2)} _\mu /g_V (0) = - 1.239 ,
\end{equation}
and
\begin{equation}
g_A (0)/g_V (0) = -1.259.
\end{equation}
$g_V$ is the vector coupling constant

\begin{equation}
g_V (q^2)/g_V (0) = 0.972,
\end{equation}
and $g_P$ is the induced pseudoscalar coupling constant,
which is less known compared to the two others. It is fixed by the
Goldhaber-Treiman relation  \cite{gotr} to be

\begin{equation}
g_P (q^2) /g_A (0) =  6.78.
\end{equation}
Ordinary calculation results are given versus this coupling constant.

\section{GENERAL FORMALISM FOR PION SCATTERING AND PION
            PHOTOPRODUCTION}

The starting point for pion scattering and pion photoproduction
calculations is the construction of the multiple scattering matrix T(E).
It is obtained as a
solution of the Lippmann-Schwinger equation

\begin{equation}
         T(E) = V(E) + V(E)\hat{P} G(E)T(E).  \label{T}
\end{equation}
Here G(E) is the pion-nucleus Green's function, V(E) is a potential
matrix, and $\hat P$ is a projection operator. Projection to the nuclear
ground state corresponds to application of the optical model, whereas
projection into a group of nuclear states corresponds to the application
of the coupled channel method.

The potential matrix is usually divided into a microscopic first order
term $V_1 (E)$ and a phenomenological second order term
$V_2 (E)$ : V(E) = $V_1$(E) + $V_2$(E).
The potential matrix $V_1 (E)$ contains the full spin and isospin
dependence of the pion nucleon
amplitudes via the impulse approximation and receives contributions
from the nuclear matrix elements.
The second order term is associated with true pion absorption and higher
order processes. As follows from Ref.\cite{E88} this term is weak
in $^6$Li. To simplify the calculations we neglect this term.

The pion photoproduction off nuclei is treated by the DWIA method. The
pion photoproduction matrix is obtained from the equation similar to
that in eq.(\ref{T}):

\begin{equation}
 T_{\pi \gamma} = U_{\pi \gamma} + T'_{\pi \pi '} G(E) U_{\pi \gamma}.
\end{equation}
Here $U_{\pi \gamma}$ is the plane wave photoproduction amplitude off
nuclei,

$E(q)$ = $E_\pi (q)$ + $E_A (q)$
is the total energy of the pion-nuclear system.
The auxiliary matrix T$'$ is related to the pion-nuclear scattering
T-matrix discussed above by the following way:

\begin{equation}
T' (\hat q , \hat q_0) = [(A-1)/A]  T(\hat q ,\hat q_o) .
\end{equation}
In the Impulse Approximation the matrix elements of $U_{\pi \gamma}$
are expressed in terms of the elementary ($\gamma$,$\pi$) amplitude.
In our calculations the Blomqwist-Laget amplitude for pion
photoproduction (see Ref.~\cite{blomq}) is used.

\newpage

\begin{table}[th]
\caption{{\em Renormalization effects for some
matrix elements in electron scattering on $^6$Li.
A --- for elastic scattering,
B --- for the transition to the $J^\pi T=3^+0$ level,
C --- for transition to the $J^\pi T=0^+1$ level}}
\vspace{5mm}
\begin{tabular}{|c|c|c|c|c|c|c|c|} \hline
\multicolumn{2}{|c|}{}&\multicolumn{2}{c}{MDM}
&\multicolumn{2}{|c|}{AMDM$_{\rm C}$}&
\multicolumn{2}{c|}{AMDM$_{\rm K}$} \\ \cline{3-8}
\multicolumn{2}{|c|}{}& $q=0.5 fm^{-1}$& $q=2.5 fm^{-1}$
& $q=0.5 fm^{-1}$& $q=2.5 fm^{-1}$
& $q=0.5 fm^{-1}$& $q=2.5 fm^{-1}$ \\ \hline
A& [000]$_0$    &.4669 &6.290E--3 &.4723 &7.981E--3 &.4657
&5.261E--3 \\ \hline
B& [022]$_0$    &3.825E--2 &1.621E--2 &3.375E--2 &1.752E--2 &3.888E--2
&1.656E--2 \\ \hline
 & [101]$_1$     &--.2004 &4.235E--3 &--.2081 &1.081E--2 &--.1954
 &8.840E--3 \\
 & [121]$_1$      &--3.839E--3 &--5.728E--4 &--4.016E--3 &--8.531E--4
 &--4.152E--2 &--7.007E--4 \\
C& [111p]$_1/M_N$ &--2.199E--3 &--4.629E--4 &--2.883E--3 &--6.968E--4
&--2.684E--3 &--6.043E--4 \\
 & [011p]$_1/M_N$ &7.303E--3 &--1.097E--3 &7.731E--3 &--1.939E--3
 &7.728E--3 &--1.546E--3 \\  \hline
\end{tabular}
\end{table}

\vspace{15mm}

\begin{table}[th]
\caption{{\em Static properties of six-nucleon nuclei}}
\vspace{5mm}
\begin{tabular}{|c|c|c|c|c|c|} \hline
Nucleus& $<r^2>^{1/2}_{ch}, fm$ & $\mu/\mu_0$ & Q, $fm^{2}$
&$<r^2>^{1/2}_{body}, fm$& $r_{halo}, fm$ \\
                           & $^6$Li      &  $^6$Li
&  $^6$Li                        &      $^6$He &  $^6$He  \\ \hline
 MDM      & 2.55 & 0.829 & 0.49 & 2.43 & 0.74\\ \hline
 AMDM$_{\rm C}$ & 2.48 & 0.838 & 0.49 & 2.33 &\\ \hline
 AMDM$_{\rm K}$ & 2.55 & 0.829 & 0.51 & 2.44 & 0.80\\ \hline
 Exper.   & 2.56 & 0.822 & -0.082& 2.33$\pm$0.04 & 0.87$\pm$0.06
                                                        \cite{tanih}
  \\
 \hline
\end{tabular}
\end{table}

\newpage

\begin{table}[th]
\caption{\em Amplitudes of muon capture in $^6$Li, their ratio and
capture rates for three versions
of the model. Numbers with $^*$ are given for results obtained without
the velocity terms.}
\vspace{5mm}
\begin{tabular}{|c|c|c|c|} \hline
& MDM & AMDM$_{\rm C}$ & AMDM$_{\rm K}$ \\ \hline
$\Delta T_1$ & 4.74E--3 & 2.47E--3 & 3.97E--3     \\
             &          &          & 1.56E--3$^*$ \\ \hline
$\Delta T_2$ & --.136   & --.139   & --.147       \\
             &          &          & 0.031$^*$    \\ \hline
$T_1$        &  0.198   & 0.205    & 0.193   \\   \hline
$T_2$        &  0.0687  & 0.0708   & 0.0660  \\   \hline
$X$          & --0.152  & --0.159  & --0.168 \\   \hline
$\Lambda$, s${-1}$
             &1259      &1351      &    1192 \\
             &          &          &1225$^*$  \\  \hline

$\Lambda_+/\Lambda_-$
             &0.0417    &0.0423    &0.0431 \\
             &          &          & 0.0247$^*$ \\ \hline
\end{tabular}
\end{table}

\newpage

\vspace{2cm}

\normalsize

\centerline{\bf Figure captions}

\vspace{1cm}

 Fig.1. The Jacoby coordinates of the $\alpha$-2N system.

 Fig.2. Longitudinal elastic form factor of $^6$Li.
        The solid line --- calculation within the framework of the
        AMDM$_{\rm K}$,
        dashed line ---  within the AMDM$_{\rm C}$ and
        dotted line --- within the MDM. Experimental data from
        Ref.\cite{GCL}.

 Fig.3. The square root of the reduced quadrupole transition strength
        B(C2,q) in units e$fm^2$ for transition to the $J^\pi T=3^+0$
        level in $^6$Li. Experimental data from Refs.\cite{Be,Fi,BDN}
        The notations are as in Fig.2.

 Fig.4. The form factor for transition to the $J^\pi T=3^+0$ level of
        $^6$Li.
        Open triangles are experimental data from Ref.\cite{BDN}, solid
        triangles are data from Ref.\cite{GCL}.
        The notations are as in Fig.2.

 Fig.5. Elastic magnetic form factor of $^6$Li. Experimental data from
        Ref.\cite{BKN}.
        The notations are as in Fig.2.

 Fig.6. Inelastic magnetic form factor for transition to the
        $J^\pi T = 0^+ 1$ level of $^6$Li. Experimental data from
        Ref.\cite{BDN}.
        The notations are as in Fig.2.

 Fig.7. Muon capture rate for transition to the ground state of $^6$He.
        Experimental data from Ref.\cite{muon6li}.
        The notations are as in Fig.2.

 Fig.8. Ratio of the hyperfine muon capture rates  to the ground
        state of $^6$He.
        The notations are as in Fig.2.

 Fig.9. The differential pion elastic cross section on $^6$Li at
        $T_\pi$ =134 MeV.
        The second order pion-nucleus potential is omitted.
        The notations are as in Fig.2.

Fig.10. The differential inelastic pion cross section to the
        $J^\pi T = 3^+ 0$ level at $T_\pi$ = 134 MeV.
        The second order pion-nucleus potential is omitted.
        The notations are as in Fig.2.

Fig.11. The vector analyzing power for pion elastic scattering on
        polarized $^6$Li at $T_\pi$ = 134 MeV.
        The second order pion-nucleus potential is omitted.
        The notations are as in Fig.2.

Fig.12. The vector analysing power for pion inelastic scattering to the
        $J^\pi T=3^+0$ level on polarized $^6$Li at $T_\pi$ =134 MeV.
        The second order pion-nucleus potential is omitted.
        The notations are as in Fig.2.

Fig.13. Experimental and calculated differential cross section for pion 
        photoproduction
        on $^6$Li at $T_\gamma = 200\ MeV$ with formation of $^6$He in
        its ground state for three versions of the model.
        The notations are as in Fig.2. Experimental data are from
        Refs. \cite{P1,P2}.

Fig.14. Calculated differential cross section for pion photoproduction
        on $^6$Li at $T_\gamma = 320\ MeV$ with formation of $^6$He in its
        ground state for three versions of the model.
        The notations are as in Fig.2.

Fig.15. Calculated beam asymmetry for pion photoproduction on $^6$Li at
        $T_\gamma = 320\ MeV$ with formation of $^6$He in its ground
        state for three versions of the model.
        The notations are as in Fig.2.

Fig.16.  Calculated beam asymmetry for pion photoproduction on $^6$Li at
         $T_\gamma = 320\ MeV$ with formation of $^6$He in its ground
         state for three values of $\Delta$ --- isobar mass in nucleus:
         solid line --- the mass is equal to the mass of the free
         particle ($M_\Delta$),
         short dashed line --- the mass is larger by 5\%
         than the free mass, long dashed line ---
         the mass is smaller by 5\% than the free mass.
         The calculation is within the framework of the AMDM$_{\rm K}$.

Fig.17. Calculated differential cross section for pion photoproduction
        on $^6$Li at $T_\gamma = 320\ MeV$ with the formation of $^6$He
        in its ground state for three values of $\Delta$ --- isobar mass
        in nucleus:
        solid line --- the mass is equal to the mass of the free
        particle ($M_\Delta$),
        short dashed line --- the mass is larger by 5\%
        than the free mass, long dashed line --- t
        he mass is smaller by 5\% than the free mass.
        The calculation is within the framework of the AMDM$_{\rm K}$.


\baselineskip=18pt
\small

\begin{thebibliography}{99}
\bibitem[1]{gfmc} B.S.Pudliner, V.R.Pandharipande, J.Carlson and
R.B.Wiringa,  Phys. Rev. Let. {\bf 74},4396 (1995)
\bibitem[2]{KK} V.I.Kukulin and V.M.Krasnopol'sky, J.Phys. {\bf G3},
795 (1977)
\bibitem[3]{svm} K.Varga and Y.Suzuki, Phys. Rev {\bf C52}, 2885 (1995)
\bibitem[4]{K-harm} G.L.Strobel. Phys. Rev. {\bf C18}, 2395 (1978)
\bibitem[5]{gorbatov}  A.M.Gorbatov, Yadernaya Fizika {\bf 55}, 1791
(1992)
\bibitem[6]{sh-mod} D.C.Zheng, B.R.Barrett, J.P.Vary, W.C.Haxton and
S.-L.Song,  Phys.Rev. {\bf C52}, 2488 (1995)
\bibitem[7]{sh2} R.F.Bishop, M.F.Flynn, M.C.Bosc\'a,
E.Buend\'{\i}a and R.Guardiola, Phys. Rev. {\bf C42}, 1341 (1990)
\bibitem[8]{1-rgm} J.A.Weeler, Phys.Rev. {\bf 52}, 1083 (1937)
\bibitem[9]{lang} K. Langanke,  Advances in Nuclear Physics
{\bf 21}, 85 (1994),  Eds. J. W. Negele and E. Vogt,
Plenum Press, New York - London
\bibitem[10]{varga1} K.Varga, Y.Suzuki and Y.Ohbayasi,
Phys.Rev. {\bf C50}, 189 (1994)
\bibitem[11]{varga2} K.Varga, Y.Suzuki and I.Tanihata,
Phys.Rev. {\bf C52}, 3013 (1995)
\bibitem[12]{wt} K.Wildermuth and Y.Tang, A unified theory of the
nucleus. Vieweg/Braunschweig, 1977
\bibitem[13]{fuji} Y.Fujiwara and Y.C.Tang,
Few-Body Systems {\bf 12}, 21 (1992)
\bibitem[14]{kruppa} A.T.Kruppa, R.Beck and F.Dickmann,
Phys. Rew. {\bf C36}, 327 (1987)
\bibitem[15]{kajino} T.Kajino, T.Matsue and A.Arima,
Nucl.Phys. {\bf A423}, 323 (1984)
\bibitem[16]{hofmann} S.Weber, M.Kachelrie\ss, M.Unkelbach and
H.M.Hofmann, Phys.Rev. {\bf C50}, 1492 (1994); \\
 M.Unkelbach and H.M.Hofmann, Few-Body Systems {\bf 11}, 143 (1991)
\bibitem[17]{efros} M.N.Ustinin and V.D.Efros Yad.Fiz.
(Sov.J.Nucl.Phys.) {\bf 49}, 1297 (1989)
\bibitem[18]{clust} V.I.Kukulin, V.G.Neudatchin, Yu.F.Smirnov and
I.T.Obukhovsky,
Clusters  as Subsystem in Light Nuclei (In the book series: Clustering
Phenomena in Nuclei, V.III. Vieweg and Sohn,
Braunschweig/Wiesbaden, 1982)
\bibitem[19]{fun92} V.I.Kukulin, V.N.Pomerantsev, Kh.D.Rasikov,
V.T.Voronchev and G.G.Ryzhikh, Nucl.Phys. {\bf A586}, 151 (1995)
\bibitem[20]{be9} V.T.Voronchev, V.I.Kukulin, V.N.Pomerantsev,
Kh.D.Rasikov and G.G.Ryzhikh,  Yad.Fizika. {\bf 57}, 1964 (1994)
\bibitem[21]{zhukov} M.V.Zhukov, B.V.Danilin, D.V.Fedorov, J.M.Bang,
      I.J.Thompson and  J.S.Vaagen, Phys.Rep., {\bf 231}, 151 (1993)
\bibitem[22]{amdm} G.G.Ryzhikh, R.A.Eramzhyan, V.I.Kukulin and
Yu.M.Tchuvil'sky,
      Nucl.Phys. {\bf A563}, 247 (1993)
\bibitem[23]{physlet} R.A.Eramzhyan, G.G.Ryzhikh, V.I.Kukulin and
Yu.M.Tchuvil'sky,  Phys.Lett. {\bf B228}, 1 (1989)
\bibitem[24]{solov} G.G.Ryzhikh, Yu.M.Tchuvil'sky and R.A.Eramzhyan,
   Frontiers in Nuclear Physics , Dubna D4-95-308, 1995, p.223
\bibitem[25]{izvest} V.I.Kukulin, G.G.Ryzhikh, R.A.Eramzhyan and
Yu.M.Tchuvil'sky,
      Izvestiya Academy Nauk USSR, ser. fiz., {\bf v.57}, 39 (1993)
\bibitem[26]{prepr} V.I.Kukulin, G.G.Ryzhikh, T.Yu.Tretiakova,
    Yu.M.Tchuvil'sky and  R.A.Eramzhyan,
    Preprints Inst. for Nucl. Res. Acad. of Sci. USSR, P-0650  and
    P-0651, Moscow (1989); P-0685, Moscow (1990)
\bibitem[27]{saito1} S.Saito, Progr. Theor.Phys (Suppl.) {\bf 62},
11 (1977)
\bibitem[28]{saito2} S.Saito, Progr. Theor. Phys {\bf 41}, 705 (1969)
\bibitem[29]{psepot} V.M.Krasnopolsky and V.I.Kukulin, Yad. Fiz.
{\bf 20}, 883 (1974)
\bibitem[30]{neud11} V.G.Neudatchin, A.A.Sakharuk, W.W.Kurowsky and
Yu.M.Tchuvil'sky, Phys.Rev. {\bf C50}, 148 (1994);
\bibitem[31]{neud12} Yu.M.Tchuvil'sky, W.W.Kurowsky, A.A.Sakharuk,
and V.G.Neudatchin, Phys.Rev. {\bf C51}, 784 (1995)
\bibitem[32]{tanih} I.Tanihata, D.Hirata, T.Kobayashi, S.Shimoura,
K.Sugimoto and H.Toki, Phys.Lett. {\bf B289}, 261 (1992)
\bibitem[33]{csoto} A.Cs\'{o}t\'{o}, Phys.Rev. {\bf C48}, 165 (1993)
\bibitem[34]{weak_cons} W.-T.Chou, E.K.Warburton and B.A.Brown,
    Phys. Rev. {\bf C47}, 163 (1993)
\bibitem[35]{GCL} G.C.Li, I.Sick, R.R.Whitney and M.R.Yearian,
Nucl.Phys. {\bf A162}, 583 (1971)
\bibitem[36]{Be} J.C.Bergstrom, Nucl.Phys. {\bf A262}, 196 (1976)
\bibitem[37]{Fi} F.Figenbrod, Zeit.Phys. {\bf 228}, 337 (1969)
\bibitem[38]{BDN} J.C.Bergstrom, U.Deutschmann and R.Neuhausen,
Nucl.Phys. {\bf A327}, 439 (1979)
\bibitem[39]{ajzenberg} F.Ajzenberg-Selove, Nucl.Phys. {\bf A490},
38 (1988)
\bibitem[40]{23.} V.I.Kukulin, V.T.Voronchev, T.D.Kaipov and
R.A.Eramzhyan, Nucl.Phys. {\bf A517}, 221 (1990)
\bibitem[41]{24.} A.Eskandarian, D.R.Lehman and W.C.Parke, Phys.Rev.
             {\bf C38}, 2341 (1988)
\bibitem[42]{BKN} J.C.Bergstrom, S.B.Kowalski and R.Neuhausen,
Phys.Rev. {\bf C25}, 1156 (1982)
\bibitem[43]{ratio} V.Brudanin e.a., Nucl. Phys.{\bf A587},
557 (1995); \\
          V.Wiaux e.a., Abstract of PANIC-96, v.1, p.488,  \\
          V.A.Kuzmin, A.A.Ovchinnikova and T.V.Tetereva,
          Yadernaya Fizika (Russian J. of Nucl.Phys.)
          {\bf 57}, 1954 (1994)  and Izvestiya RAN
          (Physics)  {\bf 59}, 163 (1995)
\bibitem[44]{muon6li} J.P.Deutsch e.a., Phys. Let. {\bf 26B},
315 (1968)
\bibitem[45]{MK} R.Mach and S.S.Kamalov, Nucl. Phys. {\bf A551},
601 (1990)
\bibitem[46]{E88} R.A.Eramzhyan , M.Gmitro, S.S.Kamalov and R.Mach,
J.Phys.G {\bf 14}, 1511 (1988); \\
R.A.Eramzhyan, M.Gmitro and S.S.Kamalov, Phys. Rev. {\bf C41},
2865 (1990)
\bibitem[47]{R94} S.Ritt et. al., Phys.Rev {\bf C52}, 2885 (1995)
\bibitem[48]{Ritt} S.Ritt et. al., Phys. Rev {\bf C43}, 745 (1991)
\bibitem[49]{P1} K.Shoda, O.Sasaki and T.Kohmura, Phys.Lett.
{\bf 101B}, 124 (1981)
\bibitem[50]{P2} J.Shaw, T.Kobayashi, W.Klayton, L.Ghedira, D.Myers,
P.Stoler, P.K.Teng, E.J.Winhold, J.H.J.Distelbrink,
Phys.Rev. {\bf C43}, 1800 (1991)
\bibitem[51]{WK} K.Wildermuth and Th.Kannelopulos, Nucl.Phys.
{\bf A7}, 150 (1958)
\bibitem[52]{TC3} Yu.M.Tchuvil'sky,
Izvestia Academy Nauk USSR, ser. fiz., {\bf 54}, 134 (1990)
\bibitem[53]{TC1} Yu.F.Smirnov and Yu.M.Tchuvils'ky, J.Phys.
{\bf G4}, L1 (1978)
\bibitem[54] {TC2} L.Ya.Glozman, V.G.Neudatchin and I.T.Obukhovsky,
Phys.Rev. {\bf C48}, 389 (1993)
\bibitem[55]{gotr} M.L.Goldberger and S.B.Treiman, Phys.Rev. {\bf 110},
1178 (1958)
\bibitem[56]{blomq} I.Blomqvist and L.M.Laget, Nucl. Phys. {\bf A280},
405 (1977); \\
 J.M.Laget, Nucl.Phys. {\bf A481}, 765 (1988)

\end{thebibliography}
\end{document}